\documentclass[twocolumn,aps,prb,superscriptaddress,floatfix]{revtex4-1}

\usepackage{xcolor}
\usepackage{graphicx}
\usepackage{dcolumn}
\usepackage{bm}
\usepackage{amsmath}

\usepackage[T1]{fontenc}
\usepackage[unicode=true,pdfusetitle,
 bookmarks=true,bookmarksnumbered=false,bookmarksopen=false,
 breaklinks=false,pdfborder={0 0 1},backref=false,colorlinks=false]
 {hyperref}
\hypersetup{
 colorlinks,linkcolor=blue,citecolor=blue,urlcolor=blue}

\newcommand{\fig}[1]{Fig.\ #1}
\newcommand{\subfig}[2]{\fig{#1}(#2)}

\begin{document}

\title{Spin Polarization driven by Itinerant Orbital Angular Momentum\\ in van der Waals Heterostructures}

\author{Luis M. Canonico}
\email{luis.canonico@icn2.cat}
\affiliation{Catalan Institute of Nanoscience and Nanotechnology (ICN2), CSIC and BIST, Campus UAB, Bellaterra, 08193 Barcelona, Spain}

\author{Jose H. Garcia}
\email{josehugo.garcia@icn2.cat}
\affiliation{Catalan Institute of Nanoscience and Nanotechnology (ICN2), CSIC and BIST, Campus UAB, Bellaterra, 08193 Barcelona, Spain}
\altaffiliation{Corresponding author. Email: josehugo.garcia@icn2.cat}

\author{Aron W. Cummings}

\affiliation{Catalan Institute of Nanoscience and Nanotechnology (ICN2), CSIC and BIST, Campus UAB, Bellaterra, 08193 Barcelona, Spain}

\author{Stephan Roche}
\affiliation{Catalan Institute of Nanoscience and Nanotechnology (ICN2), CSIC and BIST, Campus UAB, Bellaterra, 08193 Barcelona, Spain}
\affiliation{ICREA, Instituci\'o Catalana de Recerca i Estudis Avançats, 08070 Barcelona, Spain}

\date{\today}

\begin{abstract}

{We report on the possibility of manipulating magnetic materials by using itinerant orbital angular momentum to produce out-of-plane spin polarization in van der Waals heterostructures. Employing a real-space formulation of the OAM operator within linear response theory, we demonstrate that in low-symmetry transition-metal dichalcogenide (TMD) monolayers, such as 1$T{}_d$-MoTe${}_2$, the current-induced itinerant OAM exceeds the spin response by three orders of magnitude. When TMDs are coupled with ferromagnets with negligible intrinsic orbital responses, the itinerant OAM generated by the orbital Rashba-Edelstein effect transfers across the interface, generating spin densities capable of inducing magnetization dynamics inside the ferromagnet. Our findings highlight the previously overlooked role of itinerant OAM in the generation of out-of-plane spin densities, which serves as an emerging mechanism for efficient electrical control of magnetization in low-power, ultracompact storage devices.
}

\end{abstract}

\maketitle
The recent experimental confirmation of the orbital Hall effect (OHE) \cite{choi2023observation,ExpOHE2} has sparked a surge in \emph{orbitronics}, which, in analogy to spintronics, aims to develop novel mechanisms for more efficient information processing technologies using the orbital angular momentum (OAM) of the electrons. It has been shown that the magnitude of electrically induced OAM currents can surpass that of spin currents in systems with strong spin-orbit coupling (SOC), and that such currents are also significant in light, low-SOC materials at room temperature \cite{OHEBernevig,OHEorigins-1,OHEorigins-2,Go-Review}. Subsequent reports, including the inverse OHE \cite{wang2023inverse,go2024local,kashiki2025violation}, the inverse orbital Rashba-Edelstein (inverse OREE) effect \cite{el2023observation,seifert2023time}, and orbital pumping \cite{SantosOPumping,santos2024exploring,abrao2025anomalous} have shown that OAM densities generated by electrical current or magnetization dynamics can be efficiently converted into electrical signals. Furthermore, the orbital torque \cite{go2020orbital,Orbital-torque-1,Orbital-torque-2,orbital-torque-magnetic-bilayers-EXP,fukunaga2023orbital}, which refers to the induction of OAM in a ferromagnetic material to excite spin densities and control magnetization, has been confirmed in magnetic thin films and has been shown to increase the performance of magnetic nonvolatile memories \cite{gupta2025harnessing}. Notably, orbital-enhanced magnetization control has been reported to be sizable in silicon \cite{matsumoto2025observation}, highlighting the potential of this emergent degree of freedom in the development of information storage technologies.

 Theoretical modeling of OAM in solids typically follows two approaches. The first is the \emph{the atom-centered approximation} (ACA), which captures the localized orbital character of the states near the atomic sites and has been instrumental in describing orbital torques in bulk transition metal systems \cite{orbital-torque-magnetic-bilayers-EXP,go2020orbital} and uncovering orbital Hall insulating phases in two-dimensional (2D) materials such as 1H and 1T transition-metal dichalcogenides (TMDs) \cite{Us2,Us3,OHE_Bhowal_1,costa2022connecting,cysne2025orbitronicsreview}.  It also accounts for the orbital origin of the real-space localized and compensated spin textures in centrosymmetric TMDs like PtSe${}_2$ \cite{canonico2023spin}. However, the ACA becomes inadequate in systems with strong crystalline fields, where local OAM is suppressed, and in systems with structural or chemical inhomogeneities such as magnetic heterostructures \cite{PRoleBerryMoment}. In contrast, \emph{the modern theory of orbital magnetization} captures the local but also itinerant contributions of the Bloch states to the total OAM, enabling an accurate description in systems with negligible atomic orbital character. This approach is particularly relevant for 2D materials, where confinement, reduced dimensionality, and symmetry breaking can amplify orbital responses \cite{OHE_Bhowal-Vignale,nonlocalMario,TunableReversableMagnetoelectricBG,cysne2024controlling,he2020giant,serlin2020intrinsic}.

This naturally motivates the exploration of orbital phenomena in 2D systems, where high tunability, reduced symmetry, and strong interfacial effects offer unique opportunities for manipulating OAM, and raises questions about the possibility of controlling magnetization through the itinerant OAM contributions. For instance, strong out-of-plane OAM densities are expected to emerge via the OREE in monolayer and few-layer low-symmetry systems such as 1T${}_d$ WTe${}_2$ \cite{sodemann2015quantum,shi2019symmetry,ye2024nonlinear,ovalle2024orbital} or TaIrTe${}_4$ \cite{li2024room,wang2024orbital}. Such predictions are consistent with reports of unconventional out-of-plane antidamping torques in magnetic heterostructures \cite{macneill2017control,kao2022deterministic}, which enable field-free magnetization switching at room temperature \cite{zhang2023room,liu2023field,wei2023field,kajale2024field,pu2024field,wang2024field,pandey2024energy}. Although these torques have already been associated with the OREE \cite{pan2023room}, a detailed theoretical framework remains lacking. Hence, in light of these findings, it is essential to clarify the role of out-of-plane itinerant OAM in the electrical generation of spin densities in 2D magnetic van der Waals heterostructures, since understanding this mechanism is crucial for the advancement of 2D orbitronics, where the OREE may serve as the primary channel to induce orbital polarization.

In this Letter, we unveil the pivotal role of the itinerant OAM in the electrical generation of out-of-plane spin densities in 2D systems, which can be connected to the anti-damping torques reported in low-symmetry magnetic van der Waals heterostructures. We combine a real-space formulation of the OAM operator \cite{canonico2024orbital} with the kernel polynomial method \cite{Fan2021linear} and the Kubo–Bastin \cite{bastin1971quantum} formalism to study charge-to-orbital conversion in low-symmetry TMDs such as MoTe${}_2$. We show that the itinerant OAM is the dominant source of angular momentum, producing orbital densities up to three orders of magnitude larger than their spin counterparts in the isolated TMD monolayer. Furthermore, using a minimal tight-binding model of ferromagnetic materials with vanishing orbital response, we show that these low-symmetry TMDs can transfer the OAM generated by the OREE to the ferromagnetic system via proximity effects and imprint their symmetry properties. Our findings indicate that enhancing itinerant OAM contributions could accelerate the progress of magnetization control in ultracompact nanodevices.

\noindent\emph{Methodology} -- We investigate the orbital and spin susceptibility tensors within the linear-response formalism. In the static limit, the changes in the spin and orbital density due to an electric field are given by the Kubo-Bastin formula as \cite{bastin1971quantum}
\begin{align}
    \chi_{\alpha}^{\mathcal{O}} =-\frac{2\hbar e}{\Omega}\int\limits_{-\infty}^{\infty}dE F(E) \text{ImTr}\langle\mathcal{O}\partial_{E}G^{+}(E,H) v_{\alpha}\delta(E-H)\rangle,
    \label{eqn:KBformula}
\end{align}

\noindent where $\Omega$ is the volume of the sample, $\alpha$ is the direction of the applied electric field, $v_{\alpha}\equiv \frac{i}{\hbar}[H,r_{\alpha}]$ is the $\alpha$ component of the velocity operator, $r_{\alpha}$ is the position operator, $G^{\pm}(E,H)={\left[E -H \pm i0^{+}\right]}^{-1}$ is the retarded (advanced) Green's function, $F(E)=(\exp((E-\mu)/K_\mathrm{B}T)+1)^{-1}$ is the Fermi-Dirac distribution for a given temperature $T$ and chemical potential $\mu$, $\mathcal{O}=L_z(S_z)$ is the out-of-plane orbital (spin) operator, and $H$ is the Hamiltonian of the system. This formulation yields results equivalent to the eigenstate representation of the Kubo formula, which may be used to investigate orbital responses in clean systems \cite{Go-Review,OHE_Bhowal-Vignale,Us3}. However, it is also extendable to the study of real-space electrical responses in disordered materials \cite{Luis-PRL,Fan2021linear,kite}. Following the current literature, we consider the ACA representation of the orbital angular momentum ($L_z^{\text{ACA}}$) as the matrix elements of the $L_z$ operator in the atomic basis. On the other hand, to determine the itinerant contributions, we write the symmetrized OAM operator as $L_z^{\text{Mod}}=\frac{e\hbar}{4g_L\mu_\mathrm{B}}\epsilon_{\alpha\beta z}(r_{\alpha}v_{\beta}- v_{ \alpha}r_{\beta})$, where $g_L$ is the Land\'e $g$-factor for the orbital angular momentum, $\mu_\mathrm{B}$ is the Bohr magneton, and $\varepsilon_{\alpha\beta z}$ is the Levi-Civita tensor. Although several approaches have been proposed to express this operator in periodic systems \cite{OHE_Bhowal-Vignale,pezo2022orbital,liu2025quantum,cullen2025quantum}, a real-space representation of this operator, which considers the Green's function representation of the off-diagonal matrix elements of the position operator $r_{\alpha}$, has been introduced in Ref.\ \onlinecite{canonico2024orbital} to study the transport of itinerant OAM in disordered systems. It is given by

\begin{eqnarray}
 &&L_{\gamma}^{\text{Mod}} =  -\frac{\text{i}e\hbar^2\varepsilon_{\alpha\beta\gamma}}{4g_{L}\mu_B}\int dE\Big[\text{Re}(G^{+}(E,H)){v}_{\alpha}\delta(H-E){v}_{\beta}\nonumber\\
 && \hspace{25mm}+ {v}_{\alpha}\delta(H-E){v}_{\beta}\text{Re}(G^{-}(E,H))\Big],\label{eqn:RealSpaceOM}
\end{eqnarray}

\noindent where the Green functions are efficiently expanded as a series of Chebyshev polynomials \cite{canonico2024orbital} (see SM for more details \cite{suppmaterial}). This formula allows for an exact representation of the itinerant components of the OAM operator and enables the real-space study of the current-induced itinerant OAM. To study the spin and orbital susceptibilities in low-symmetry TMDs on equal footing, we will use the minimal model to describe these systems introduced in Ref.\ \onlinecite{vila2021low},

\begin{align}
  H=\sum_{i,s}(\Delta + 4m_d +\delta)c^{\dagger}_{i,s}c_{i,s} + \sum_{\langle i j\rangle,s}-(m_p+m_d)c^{\dagger}_{i,s}c_{j,s}\nonumber\\
  +\sum_{i,s}(\Delta - 4m_d -\delta)d^{\dagger}_{i,s}d_{i,s}+\sum_{\langle i j\rangle,s}-(m_p-m_d)d^{\dagger}_{i,s}d_{j,s}\nonumber\\
  +\sum_{\langle ij\rangle,ss^{\prime}}\frac{-\text{i}}{2}(\Lambda_{ss^{\prime}}\times\hat{\mathfrak{l}}_{ij})\cdot(\hat{y}+\hat{z})c^{\dagger}_{i,s}d_{j,s^{\prime}} +\sum_{is}\eta c^{\dagger}_{is}d_{is}\nonumber\\
  -\sum_{\langle ij\rangle,s} \frac{\beta}{2}(\hat{\mathfrak{l}}_{ij}\cdot\hat{y})c^{\dagger}_{i,s}d_{j,s} +\text{h.c.}\label{eqn:tmd}
   \nonumber\\
\end{align}

This effective four-band Hamiltonian considers two orbitals (plus spin) per unit cell on a rectangular lattice, $p_y$ and $d_{yz}$, where the former and the latter correspond to the chalcogen and the transition metal atoms, respectively. The operators $c_{i,s}$ and $d_{i,s}$ are the annihilation operators of each orbital at site $i$ with spin $s$. The first four terms describe the spin-degenerate valence and conduction bands with hopping amplitudes $m_p \pm m_d$, $\delta$ parametrizes the degree of band inversion at $\Gamma$, and a constant $\Delta$ is set to match the positions of the conduction band and the Fermi level with those obtained from DFT. The fifth term embodies the SOC, where $\Lambda\equiv\left(\Lambda_x\sigma_x,-\Lambda_y\sigma_y,\Lambda_z\sigma_z\right)$, $\sigma_{i}$ are the Pauli matrices in spin space, and $\hat{y}/\hat{z}$ are the unit directional vectors. The sixth term breaks inversion symmetry and can encapsulate either the effects of distortions or a perpendicular displacement field, and the final term accounts for the $x-y$ crystalline anisotropy, with $\hat{\mathfrak{l}}_{ij}$ a unit vector pointing from the site $i$ to $j$.\\

\begin{figure}[h]
\centering
\includegraphics[width=0.9\linewidth]{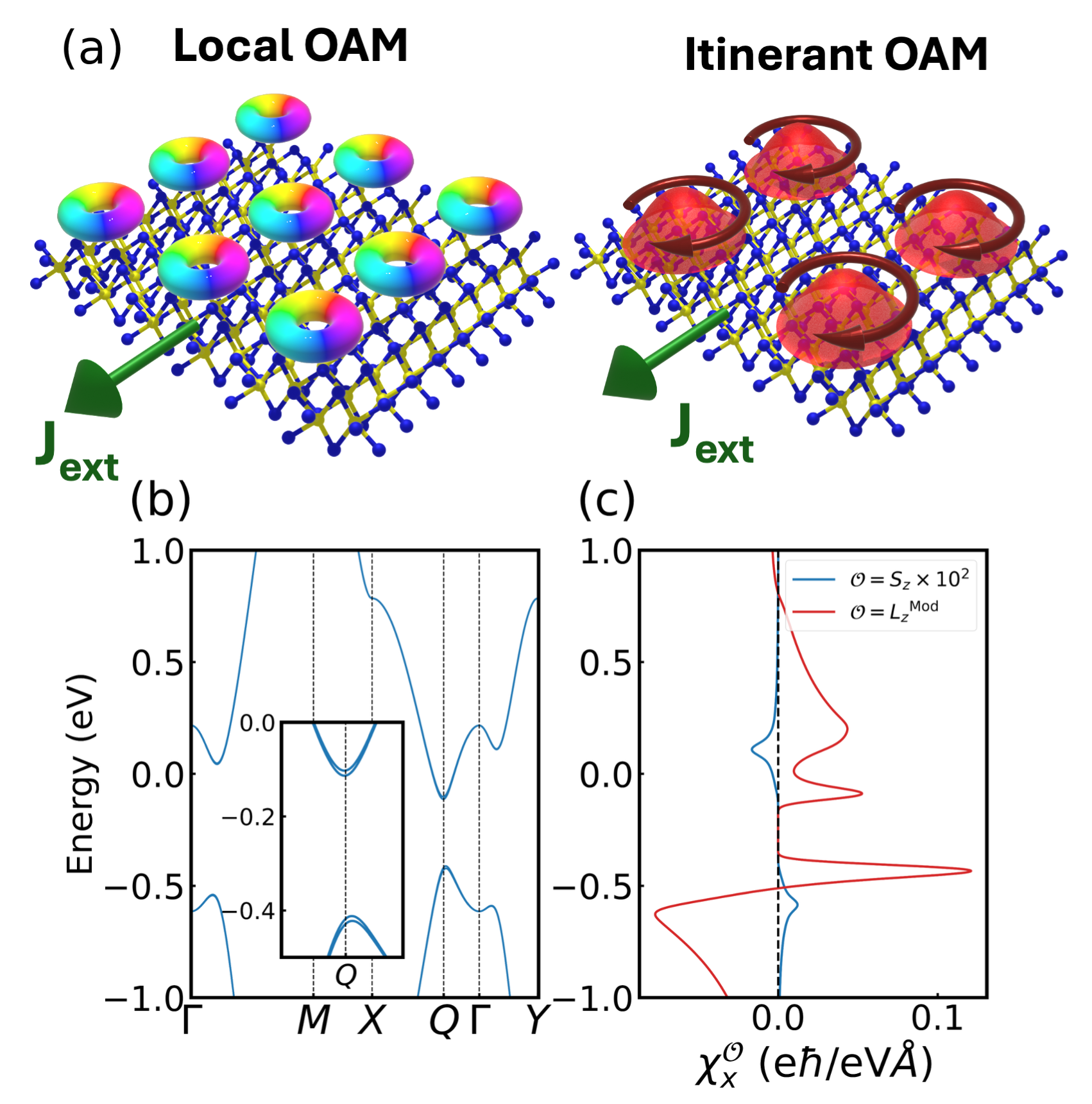}
\caption{(a) Schematic representation of the local and itinerant contributions to the orbital moment. (b) Band structure of MoTe${}_2$ model of Eq.\ \eqref{eqn:tmd}. (c) Comparison of the current-induced orbital (red) and spin (blue) densities for symmetry breaking strength $\eta = 5.4$ meV, highlighting the dominance of the itinerant orbital effect.}\label{fig:1}
\end{figure}

\noindent\emph{Orbital Rashba-Edelstein Effect in 1T${}_d$ TMDs} -- Since matrix elements of the $L_z$ operator within the ACA representation are zero for the orbitals in the model described by Eq.\ \eqref{eqn:tmd} ($p_y$ and $d_{yz}$), it implies that these orbitals cannot form a linear combination with non-zero atomic-like OAM, and hence the itinerant OAM will dominate the charge-to-orbital angular momentum conversion in this model. In \subfig{\ref{fig:1}}{b}, we show the band structure of the minimal model with tight-binding parameters for MoTe${}_2$, as a representative of the low-symmetry TMDs. A key feature of this model is the asymmetry of the energy bands along the high-symmetry paths $X-\Gamma$ and $\Gamma-Y$, characteristic of these systems. Moreover, $\eta \neq 0$ leads to band splitting at $Q$, with the bands polarized in spin and orbital angular momentum \cite{suppmaterial}.

In \subfig{\ref{fig:1}}{c}, we present the spin and orbital susceptibilities as a function of energy for an electric field applied along $x$ (note that there is no response for an electric field along $y$). Remarkably, the itinerant orbital susceptibility is nearly \emph{three orders of magnitude} larger than the spin susceptibility, highlighting the dominant role of itinerant OAM in these low-symmetry systems. As observed experimentally, the Berry curvature dipole can be reversed by applying a vertical displacement field controlled by top and bottom gates \cite{xu2018electrically}. Phenomenologically, the inversion of the nonlinear Hall effect is related to the reversal of the Berry curvature dipole \cite{sodemann2015quantum}, which in turn suggests an inversion of the charge-to-orbital conversion in the system \cite{ovalle2024orbital}. In our model, this reversal of the displacement field is effectively captured by changing the sign of $\eta$ in Eq.\ \eqref{eqn:tmd}. As shown in the Supplementary Material\cite{suppmaterial}, inverting $\eta$ leads to a sign reversal of both spin and orbital susceptibilities, while doubling $\eta$ results in a proportional enhancement of the spin and orbital responses.

\emph{Orbital-driven spin polarization in van der Waals heterostructures} -- After quantifying the predominance of orbital signals in the charge-to-angular momentum conversion in low-symmetry TMDs, we now explore whether this OAM produced in the TMD can be imprinted onto an adjacent ferromagnetic material through proximity interactions. For this, we extend the tight-binding Hamiltonian of Eq.\ \eqref{fig:1} to include a ferromagnetic (FM) material coupled via proximity to the TMD. The FM is modeled with a Slater-Koster parametrization of $p_x$, $p_y$, and $p_z$ orbitals on a square lattice, plus an exchange splitting along the $z$-direction. The reason behind the selection of this model is that a square lattice exhibits neither sizable OREE nor OHE due to symmetry restrictions and the lack of hybridization between the $p_x$, $p_y$ and $p_z$ orbitals, which is usually mediated by $s$ orbitals in this kind of system \cite{KMETellurium,OrbitalTexture}. The spin susceptibility of this model is also zero. Thus, any spin or orbital density induced in the FM should arise solely from hybridization with the TMD. To calculate the angular momentum induced in the FM by the TMD, we consider the spin ($S_z$) and OAM operators (${L_z}^{\text{ACA}}$ and ${L_{z}}^{\text{Mod}}$) projected on the FM sites \cite{suppmaterial}.

We focus on two conditions for the heterostructure interface. First, we analyze the spin and orbital susceptibilities produced in the FM when hybridization between the FM and TMD states occurs far from the pocket $Q$. As the itinerant OAM in the TMD is maximized near $Q$ \cite{suppmaterial}, this represents a ``weak'' hybridization condition. Second, we consider the case where the energy bands of the TMD and the FM hybridize near the edge of the TMD bands (in the vicinity of $Q$), representing ``strong'' hybridization.

\begin{figure*}[ht]
\centering
\includegraphics[width=0.9\linewidth]{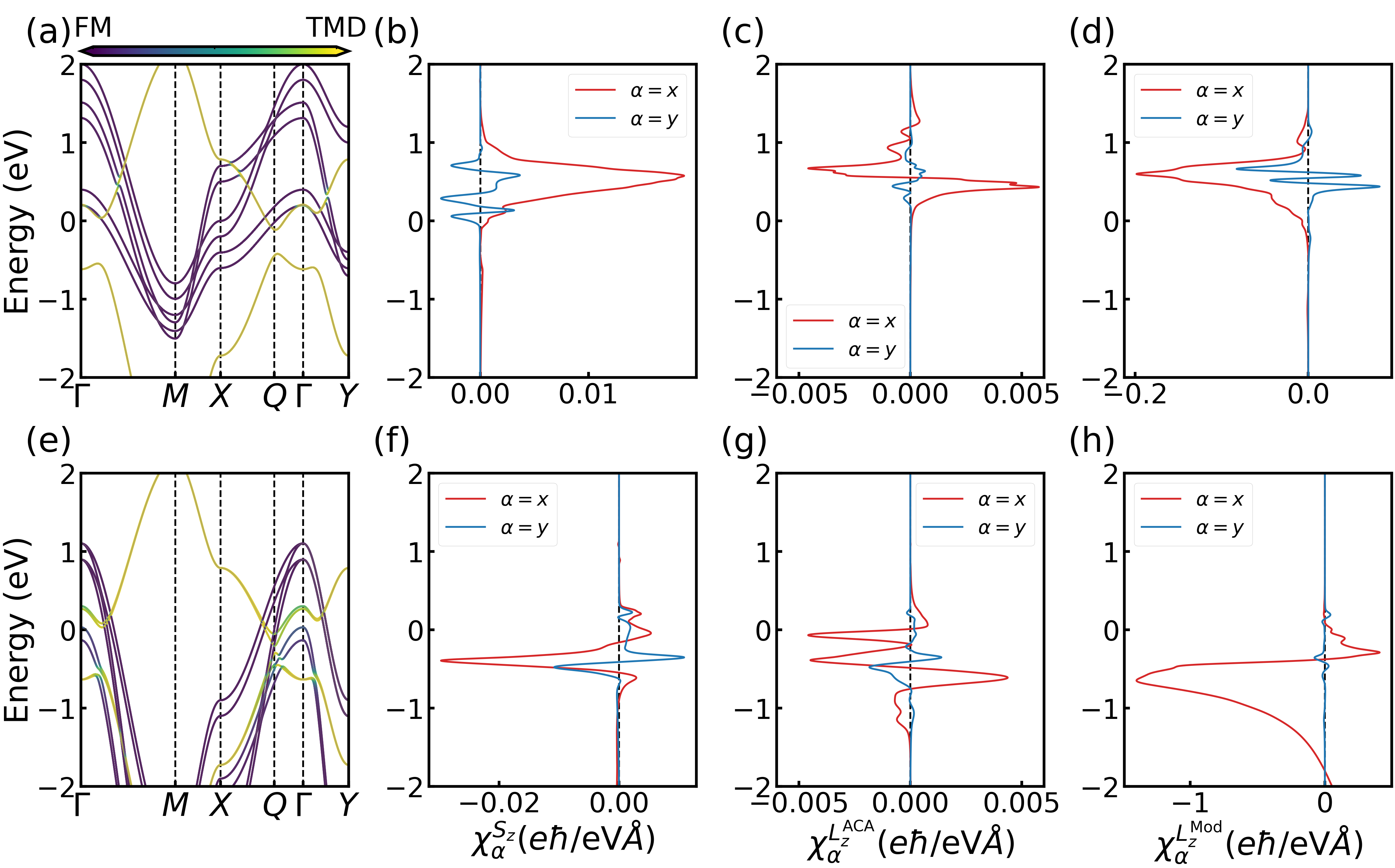}

\caption{
        (a), (e) Band structure of the MoTe$_2$/FM heterostructure for weak hybridization (away from $Q$) with $\eta=5.4$ meV (a) and strong hybridization with $\eta = 108$~meV (e). The color scale represents the real-space layer projection of the eigenstates. (b)-(d) Energy-resolved responses for weak hybridization and $\eta = 5.4$~meV: (b) spin susceptibility $\chi_{\eta}^{S_z}$, (c) atomic (ACA) orbital susceptibility $\chi_{\alpha}^{{L_z}^\text{ACA}}$, and (d) itinerant orbital susceptibility $\chi_{\alpha}^{{L_z}^\text{Mod}}$, for electric fields applied along $\alpha = x$ (red) and $\alpha = y$ (blue). 
        (f)-(h) Same responses computed for strong hybridization with $\eta = 108$~meV.}\label{fig:3}
\end{figure*}

In \subfig{\ref{fig:3}}{a} we show the energy bands of the weakly hybridized TMD/FM heterostructure, for $\eta=5.4$ meV, with their real-space projection indicated by the color scale. Here, hybridization between the FM and the TMD occurs primarily in the conduction band. The spin susceptibility is shown in \subfig{\ref{fig:3}}{b}. Comparing to the band structure, we can see that the peaks in the spin susceptibility coincide with the energy where the hybridization between the two materials occurs. However, in contrast with the case of the isolated MoTe${}_2$ monolayer, we find a nonzero current-induced spin susceptibility when the field is applied along $y$. This response can be attributed to Fermi sea contributions to the spin and orbital susceptibilities that become accessible due to the exchange splitting in the FM \cite{medina2024emerging} (See SM \cite{suppmaterial}), but is nearly one order of magnitude smaller than the response to the field along $x$. Panels (c) and (d) show the orbital response in the FM arising from the ACA and the itinerant contributions, respectively, with the latter dominating over the former.

These results show that nonzero spin and orbital responses can be imprinted into an FM via proximity with a low-symmetry TMD. This raises a fundamental question for this heterostructure: \emph{Is the current-induced spin density driven by ACA or by itinerant OAM?} The results in Fig.\ \ref{fig:3} demonstrate that the itinerant OAM is $\sim$$40\times$ larger than the localized one, suggesting the latter. Furthermore, the usual orbital torque mechanism reported from experiments \cite{orbital-torque-magnetic-bilayers-EXP} is that orbital excitations are converted into spin signals via the spin-orbit coupling -- with this conversion, the spin density should be smaller than the orbital density. In Fig.\ \ref{fig:3}, the fact that the spin density is $\sim4\times$ larger than the ACA orbital density suggests that this component does not play a significant role. 

To further inquire into the role of the itinerant OAM, we increased the crystal field splitting, modified the Slater-Koster hopping integrals of the FM, and increased $\eta$ to maximize the hybridization between the two subsystems near the band gap of the TMD. Figure \ref{fig:3}(e) show sthe real-space projection of the energy bands for $\eta=108$ meV. In contrast to \subfig{\ref{fig:3}}{a}, the bands are now strongly hybridized around the valence band maximum and conduction band minimum of the TMD.

Figure \ref{fig:3}(f) shows the spin susceptibility in the strongly hybridized system. Interestingly, the directional anisotropy of the spin susceptibility is reduced. To gain further insight into the origin of the spin susceptibility induced in the FM, as in the previous case, we compare the orbital susceptibilities computed from the ACA and the itinerant OAM. The orbital density arising from the ACA approximation, shown in \subfig{\ref{fig:3}}{g}, is $\sim5\times$ smaller than the spin density, ruling out this contribution via the usual orbital-torque mechanism. Meanwhile, the orbital density arising from the itinerant contribution is enhanced due to the vicinity to the $Q$ point. Overall, our results suggest that itinerant contributions to the orbital moment predominantly govern the current-induced orbital response in the heterostructure. In the SM \cite{suppmaterial}, we present further analyses on other FM models, modulation of the current-induced spin and orbital densities with $\eta$, and their dependence on SOC and hybridization with FM.

\emph{Conclusion} -- 
Combining a real-space formulation of the OAM operator with tight-binding models for low-symmetry TMDs and FMs, along with large-scale quantum transport simulations, we demonstrate that itinerant OAM is the dominant source of electrically generated angular momentum in low-symmetry TMDs. When interfaced with ferromagnets that exhibit negligible intrinsic spin and orbital responses, this itinerant OAM enables the efficient generation of out-of-plane spin densities in the FM. Notably, when the energy bands of the TMD and ferromagnet align near the $Q$ point where itinerant orbital responses peak, both charge-to-orbital and charge-to-spin conversion in the ferromagnet are significantly enhanced and tunable via electrostatic gating. Our findings provide evidence of the link between itinerant orbital effects and the emergence of out-of-plane spin densities that give rise to anti-damping spin-orbit torques observed in recent experiments. This study highlights the potential of itinerant orbital angular momentum for magnetization control in electronic devices, creating new avenues for the development of nonvolatile magnetic storage technologies through the engineering of itinerant OAM and, simultaneously, reducing the reliance on heavy metals in spin-orbitronic applications.

\begin{acknowledgments} 

J.H.G acknowledges funding from the ERC grant AI4SPIN, by the European Union’s Horizon Europe research and innovation programme under grant agreement No. 101078370. S.R and J.H.G, acknowledge funding from the FLAG-ERA grant MNEMOSYN, by MCIN/AEI /10.13039/501100011033 and European Union "NextGenerationEU/PRTR”  under grant PCI2021-122035-2A-2a and funding from the European Union’s Horizon 2020 research and innovation programme under grant agreement No 881603. ICN2 is funded by the CERCA Programme/Generalitat de Catalunya and supported by the Severo Ochoa Centres of Excellence programme, Grant CEX2021-001214-S, funded by MCIN/AEI/10.13039.501100011033.  This work is also supported by MICIN with European funds‐NextGenerationEU (PRTR‐C17.I1) and by 2021 SGR 00997, funded by Generalitat de Catalunya.
\end{acknowledgments}


\begin{thebibliography}{64}%
\makeatletter
\providecommand \@ifxundefined [1]{%
 \@ifx{#1\undefined}
}%
\providecommand \@ifnum [1]{%
 \ifnum #1\expandafter \@firstoftwo
 \else \expandafter \@secondoftwo
 \fi
}%
\providecommand \@ifx [1]{%
 \ifx #1\expandafter \@firstoftwo
 \else \expandafter \@secondoftwo
 \fi
}%
\providecommand \natexlab [1]{#1}%
\providecommand \enquote  [1]{``#1''}%
\providecommand \bibnamefont  [1]{#1}%
\providecommand \bibfnamefont [1]{#1}%
\providecommand \citenamefont [1]{#1}%
\providecommand \href@noop [0]{\@secondoftwo}%
\providecommand \href [0]{\begingroup \@sanitize@url \@href}%
\providecommand \@href[1]{\@@startlink{#1}\@@href}%
\providecommand \@@href[1]{\endgroup#1\@@endlink}%
\providecommand \@sanitize@url [0]{\catcode `\\12\catcode `\$12\catcode
  `\&12\catcode `\#12\catcode `\^12\catcode `\_12\catcode `\%12\relax}%
\providecommand \@@startlink[1]{}%
\providecommand \@@endlink[0]{}%
\providecommand \url  [0]{\begingroup\@sanitize@url \@url }%
\providecommand \@url [1]{\endgroup\@href {#1}{\urlprefix }}%
\providecommand \urlprefix  [0]{URL }%
\providecommand \Eprint [0]{\href }%
\providecommand \doibase [0]{http://dx.doi.org/}%
\providecommand \selectlanguage [0]{\@gobble}%
\providecommand \bibinfo  [0]{\@secondoftwo}%
\providecommand \bibfield  [0]{\@secondoftwo}%
\providecommand \translation [1]{[#1]}%
\providecommand \BibitemOpen [0]{}%
\providecommand \bibitemStop [0]{}%
\providecommand \bibitemNoStop [0]{.\EOS\space}%
\providecommand \EOS [0]{\spacefactor3000\relax}%
\providecommand \BibitemShut  [1]{\csname bibitem#1\endcsname}%
\let\auto@bib@innerbib\@empty
\bibitem [{\citenamefont {Choi}\ \emph {et~al.}(2023)\citenamefont {Choi},
  \citenamefont {Jo}, \citenamefont {Ko}, \citenamefont {Go}, \citenamefont
  {Kim}, \citenamefont {Park}, \citenamefont {Kim}, \citenamefont {Min},
  \citenamefont {Choi},\ and\ \citenamefont {Lee}}]{choi2023observation}%
  \BibitemOpen
  \bibfield  {author} {\bibinfo {author} {\bibfnamefont {Y.-G.}\ \bibnamefont
  {Choi}}, \bibinfo {author} {\bibfnamefont {D.}~\bibnamefont {Jo}}, \bibinfo
  {author} {\bibfnamefont {K.-H.}\ \bibnamefont {Ko}}, \bibinfo {author}
  {\bibfnamefont {D.}~\bibnamefont {Go}}, \bibinfo {author} {\bibfnamefont
  {K.-H.}\ \bibnamefont {Kim}}, \bibinfo {author} {\bibfnamefont {H.~G.}\
  \bibnamefont {Park}}, \bibinfo {author} {\bibfnamefont {C.}~\bibnamefont
  {Kim}}, \bibinfo {author} {\bibfnamefont {B.-C.}\ \bibnamefont {Min}},
  \bibinfo {author} {\bibfnamefont {G.-M.}\ \bibnamefont {Choi}}, \ and\
  \bibinfo {author} {\bibfnamefont {H.-W.}\ \bibnamefont {Lee}},\ }\href@noop
  {} {\bibfield  {journal} {\bibinfo  {journal} {Nature}\ }\textbf {\bibinfo
  {volume} {619}},\ \bibinfo {pages} {52} (\bibinfo {year} {2023})}\BibitemShut
  {NoStop}%
\bibitem [{\citenamefont {Lyalin}\ \emph {et~al.}(2023)\citenamefont {Lyalin},
  \citenamefont {Alikhah}, \citenamefont {Berritta}, \citenamefont {Oppeneer},\
  and\ \citenamefont {Kawakami}}]{ExpOHE2}%
  \BibitemOpen
  \bibfield  {author} {\bibinfo {author} {\bibfnamefont {I.}~\bibnamefont
  {Lyalin}}, \bibinfo {author} {\bibfnamefont {S.}~\bibnamefont {Alikhah}},
  \bibinfo {author} {\bibfnamefont {M.}~\bibnamefont {Berritta}}, \bibinfo
  {author} {\bibfnamefont {P.~M.}\ \bibnamefont {Oppeneer}}, \ and\ \bibinfo
  {author} {\bibfnamefont {R.~K.}\ \bibnamefont {Kawakami}},\ }\href {\doibase
  10.1103/PhysRevLett.131.156702} {\bibfield  {journal} {\bibinfo  {journal}
  {Phys. Rev. Lett.}\ }\textbf {\bibinfo {volume} {131}},\ \bibinfo {pages}
  {156702} (\bibinfo {year} {2023})}\BibitemShut {NoStop}%
\bibitem [{\citenamefont {Bernevig}\ \emph {et~al.}(2005)\citenamefont
  {Bernevig}, \citenamefont {Hughes},\ and\ \citenamefont
  {Zhang}}]{OHEBernevig}%
  \BibitemOpen
  \bibfield  {author} {\bibinfo {author} {\bibfnamefont {B.~A.}\ \bibnamefont
  {Bernevig}}, \bibinfo {author} {\bibfnamefont {T.~L.}\ \bibnamefont
  {Hughes}}, \ and\ \bibinfo {author} {\bibfnamefont {S.-C.}\ \bibnamefont
  {Zhang}},\ }\href {\doibase 10.1103/PhysRevLett.95.066601} {\bibfield
  {journal} {\bibinfo  {journal} {Phys. Rev. Lett.}\ }\textbf {\bibinfo
  {volume} {95}},\ \bibinfo {pages} {066601} (\bibinfo {year}
  {2005})}\BibitemShut {NoStop}%
\bibitem [{\citenamefont {Kontani}\ \emph {et~al.}(2008)\citenamefont
  {Kontani}, \citenamefont {Tanaka}, \citenamefont {Hirashima}, \citenamefont
  {Yamada},\ and\ \citenamefont {Inoue}}]{OHEorigins-1}%
  \BibitemOpen
  \bibfield  {author} {\bibinfo {author} {\bibfnamefont {H.}~\bibnamefont
  {Kontani}}, \bibinfo {author} {\bibfnamefont {T.}~\bibnamefont {Tanaka}},
  \bibinfo {author} {\bibfnamefont {D.~S.}\ \bibnamefont {Hirashima}}, \bibinfo
  {author} {\bibfnamefont {K.}~\bibnamefont {Yamada}}, \ and\ \bibinfo {author}
  {\bibfnamefont {J.}~\bibnamefont {Inoue}},\ }\href {\doibase
  10.1103/PhysRevLett.100.096601} {\bibfield  {journal} {\bibinfo  {journal}
  {Phys. Rev. Lett.}\ }\textbf {\bibinfo {volume} {100}},\ \bibinfo {pages}
  {096601} (\bibinfo {year} {2008})}\BibitemShut {NoStop}%
\bibitem [{\citenamefont {Tanaka}\ \emph {et~al.}(2008)\citenamefont {Tanaka},
  \citenamefont {Kontani}, \citenamefont {Naito}, \citenamefont {Naito},
  \citenamefont {Hirashima}, \citenamefont {Yamada},\ and\ \citenamefont
  {Inoue}}]{OHEorigins-2}%
  \BibitemOpen
  \bibfield  {author} {\bibinfo {author} {\bibfnamefont {T.}~\bibnamefont
  {Tanaka}}, \bibinfo {author} {\bibfnamefont {H.}~\bibnamefont {Kontani}},
  \bibinfo {author} {\bibfnamefont {M.}~\bibnamefont {Naito}}, \bibinfo
  {author} {\bibfnamefont {T.}~\bibnamefont {Naito}}, \bibinfo {author}
  {\bibfnamefont {D.~S.}\ \bibnamefont {Hirashima}}, \bibinfo {author}
  {\bibfnamefont {K.}~\bibnamefont {Yamada}}, \ and\ \bibinfo {author}
  {\bibfnamefont {J.}~\bibnamefont {Inoue}},\ }\href {\doibase
  10.1103/PhysRevB.77.165117} {\bibfield  {journal} {\bibinfo  {journal} {Phys.
  Rev. B}\ }\textbf {\bibinfo {volume} {77}},\ \bibinfo {pages} {165117}
  (\bibinfo {year} {2008})}\BibitemShut {NoStop}%
\bibitem [{\citenamefont {Go}\ \emph {et~al.}(2021)\citenamefont {Go},
  \citenamefont {Jo}, \citenamefont {Lee}, \citenamefont {Kläui},\ and\
  \citenamefont {Mokrousov}}]{Go-Review}%
  \BibitemOpen
  \bibfield  {author} {\bibinfo {author} {\bibfnamefont {D.}~\bibnamefont
  {Go}}, \bibinfo {author} {\bibfnamefont {D.}~\bibnamefont {Jo}}, \bibinfo
  {author} {\bibfnamefont {H.-W.}\ \bibnamefont {Lee}}, \bibinfo {author}
  {\bibfnamefont {M.}~\bibnamefont {Kläui}}, \ and\ \bibinfo {author}
  {\bibfnamefont {Y.}~\bibnamefont {Mokrousov}},\ }\href {\doibase
  10.1209/0295-5075/ac2653} {\bibfield  {journal} {\bibinfo  {journal} {{EPL}
  (Europhysics Letters)}\ }\textbf {\bibinfo {volume} {135}},\ \bibinfo {pages}
  {37001} (\bibinfo {year} {2021})}\BibitemShut {NoStop}%
\bibitem [{\citenamefont {Wang}\ \emph {et~al.}(2023)\citenamefont {Wang},
  \citenamefont {Feng}, \citenamefont {Yang}, \citenamefont {Zhang},
  \citenamefont {Liu}, \citenamefont {Xu}, \citenamefont {Jia}, \citenamefont
  {Wu}, \citenamefont {Yu}, \citenamefont {Xu} \emph
  {et~al.}}]{wang2023inverse}%
  \BibitemOpen
  \bibfield  {author} {\bibinfo {author} {\bibfnamefont {P.}~\bibnamefont
  {Wang}}, \bibinfo {author} {\bibfnamefont {Z.}~\bibnamefont {Feng}}, \bibinfo
  {author} {\bibfnamefont {Y.}~\bibnamefont {Yang}}, \bibinfo {author}
  {\bibfnamefont {D.}~\bibnamefont {Zhang}}, \bibinfo {author} {\bibfnamefont
  {Q.}~\bibnamefont {Liu}}, \bibinfo {author} {\bibfnamefont {Z.}~\bibnamefont
  {Xu}}, \bibinfo {author} {\bibfnamefont {Z.}~\bibnamefont {Jia}}, \bibinfo
  {author} {\bibfnamefont {Y.}~\bibnamefont {Wu}}, \bibinfo {author}
  {\bibfnamefont {G.}~\bibnamefont {Yu}}, \bibinfo {author} {\bibfnamefont
  {X.}~\bibnamefont {Xu}},  \emph {et~al.},\ }\href@noop {} {\bibfield
  {journal} {\bibinfo  {journal} {npj Quantum Materials}\ }\textbf {\bibinfo
  {volume} {8}},\ \bibinfo {pages} {28} (\bibinfo {year} {2023})}\BibitemShut
  {NoStop}%
\bibitem [{\citenamefont {Go}\ \emph {et~al.}()\citenamefont {Go},
  \citenamefont {Seifert}, \citenamefont {Kampfrath}, \citenamefont {Ando},
  \citenamefont {Lee},\ and\ \citenamefont {Mokrousov}}]{go2024local}%
  \BibitemOpen
  \bibfield  {author} {\bibinfo {author} {\bibfnamefont {D.}~\bibnamefont
  {Go}}, \bibinfo {author} {\bibfnamefont {T.~S.}\ \bibnamefont {Seifert}},
  \bibinfo {author} {\bibfnamefont {T.}~\bibnamefont {Kampfrath}}, \bibinfo
  {author} {\bibfnamefont {K.}~\bibnamefont {Ando}}, \bibinfo {author}
  {\bibfnamefont {H.-W.}\ \bibnamefont {Lee}}, \ and\ \bibinfo {author}
  {\bibfnamefont {Y.}~\bibnamefont {Mokrousov}},\ }\href@noop {} {\bibinfo
  {journal} {arXiv:2407.00517}\ }\BibitemShut {NoStop}%
\bibitem [{\citenamefont {Kashiki}\ \emph {et~al.}()\citenamefont {Kashiki},
  \citenamefont {Hayashi}, \citenamefont {Go}, \citenamefont {Mokrousov},\ and\
  \citenamefont {Ando}}]{kashiki2025violation}%
  \BibitemOpen
\bibfield  {journal} {  }\bibfield  {author} {\bibinfo {author} {\bibfnamefont
  {H.}~\bibnamefont {Kashiki}}, \bibinfo {author} {\bibfnamefont
  {H.}~\bibnamefont {Hayashi}}, \bibinfo {author} {\bibfnamefont
  {D.}~\bibnamefont {Go}}, \bibinfo {author} {\bibfnamefont {Y.}~\bibnamefont
  {Mokrousov}}, \ and\ \bibinfo {author} {\bibfnamefont {K.}~\bibnamefont
  {Ando}},\ }\href@noop {} {\bibinfo  {journal} {arXiv:2504.05139}\
  }\BibitemShut {NoStop}%
\bibitem [{\citenamefont {El~Hamdi}\ \emph {et~al.}(2023)\citenamefont
  {El~Hamdi}, \citenamefont {Chauleau}, \citenamefont {Boselli}, \citenamefont
  {Thibault}, \citenamefont {Gorini}, \citenamefont {Smogunov}, \citenamefont
  {Barreteau}, \citenamefont {Gariglio}, \citenamefont {Triscone},\ and\
  \citenamefont {Viret}}]{el2023observation}%
  \BibitemOpen
\bibfield  {journal} {  }\bibfield  {author} {\bibinfo {author} {\bibfnamefont
  {A.}~\bibnamefont {El~Hamdi}}, \bibinfo {author} {\bibfnamefont {J.-Y.}\
  \bibnamefont {Chauleau}}, \bibinfo {author} {\bibfnamefont {M.}~\bibnamefont
  {Boselli}}, \bibinfo {author} {\bibfnamefont {C.}~\bibnamefont {Thibault}},
  \bibinfo {author} {\bibfnamefont {C.}~\bibnamefont {Gorini}}, \bibinfo
  {author} {\bibfnamefont {A.}~\bibnamefont {Smogunov}}, \bibinfo {author}
  {\bibfnamefont {C.}~\bibnamefont {Barreteau}}, \bibinfo {author}
  {\bibfnamefont {S.}~\bibnamefont {Gariglio}}, \bibinfo {author}
  {\bibfnamefont {J.-M.}\ \bibnamefont {Triscone}}, \ and\ \bibinfo {author}
  {\bibfnamefont {M.}~\bibnamefont {Viret}},\ }\href@noop {} {\bibfield
  {journal} {\bibinfo  {journal} {Nature Physics}\ }\textbf {\bibinfo {volume}
  {19}},\ \bibinfo {pages} {1855} (\bibinfo {year} {2023})}\BibitemShut
  {NoStop}%
\bibitem [{\citenamefont {Seifert}\ \emph {et~al.}(2023)\citenamefont
  {Seifert}, \citenamefont {Go}, \citenamefont {Hayashi}, \citenamefont
  {Rouzegar}, \citenamefont {Freimuth}, \citenamefont {Ando}, \citenamefont
  {Mokrousov},\ and\ \citenamefont {Kampfrath}}]{seifert2023time}%
  \BibitemOpen
  \bibfield  {author} {\bibinfo {author} {\bibfnamefont {T.~S.}\ \bibnamefont
  {Seifert}}, \bibinfo {author} {\bibfnamefont {D.}~\bibnamefont {Go}},
  \bibinfo {author} {\bibfnamefont {H.}~\bibnamefont {Hayashi}}, \bibinfo
  {author} {\bibfnamefont {R.}~\bibnamefont {Rouzegar}}, \bibinfo {author}
  {\bibfnamefont {F.}~\bibnamefont {Freimuth}}, \bibinfo {author}
  {\bibfnamefont {K.}~\bibnamefont {Ando}}, \bibinfo {author} {\bibfnamefont
  {Y.}~\bibnamefont {Mokrousov}}, \ and\ \bibinfo {author} {\bibfnamefont
  {T.}~\bibnamefont {Kampfrath}},\ }\href {\doibase 10.1038/s41565-023-01470-8}
  {\bibfield  {journal} {\bibinfo  {journal} {Nature Nanotechnology}\ }\textbf
  {\bibinfo {volume} {18}},\ \bibinfo {pages} {1132} (\bibinfo {year}
  {2023})}\BibitemShut {NoStop}%
\bibitem [{\citenamefont {Santos}\ \emph {et~al.}(2023)\citenamefont {Santos},
  \citenamefont {Abr\~ao}, \citenamefont {Go}, \citenamefont {de~Assis},
  \citenamefont {Mokrousov}, \citenamefont {Mendes},\ and\ \citenamefont
  {Azevedo}}]{SantosOPumping}%
  \BibitemOpen
  \bibfield  {author} {\bibinfo {author} {\bibfnamefont {E.}~\bibnamefont
  {Santos}}, \bibinfo {author} {\bibfnamefont {J.}~\bibnamefont {Abr\~ao}},
  \bibinfo {author} {\bibfnamefont {D.}~\bibnamefont {Go}}, \bibinfo {author}
  {\bibfnamefont {L.}~\bibnamefont {de~Assis}}, \bibinfo {author}
  {\bibfnamefont {Y.}~\bibnamefont {Mokrousov}}, \bibinfo {author}
  {\bibfnamefont {J.}~\bibnamefont {Mendes}}, \ and\ \bibinfo {author}
  {\bibfnamefont {A.}~\bibnamefont {Azevedo}},\ }\href {\doibase
  10.1103/PhysRevApplied.19.014069} {\bibfield  {journal} {\bibinfo  {journal}
  {Phys. Rev. Appl.}\ }\textbf {\bibinfo {volume} {19}},\ \bibinfo {pages}
  {014069} (\bibinfo {year} {2023})}\BibitemShut {NoStop}%
\bibitem [{\citenamefont {Santos}\ \emph {et~al.}(2024)\citenamefont {Santos},
  \citenamefont {Abr{\~a}o}, \citenamefont {Vieira}, \citenamefont {Mendes},
  \citenamefont {Rodr{\'\i}guez-Su{\'a}rez},\ and\ \citenamefont
  {Azevedo}}]{santos2024exploring}%
  \BibitemOpen
  \bibfield  {author} {\bibinfo {author} {\bibfnamefont {E.}~\bibnamefont
  {Santos}}, \bibinfo {author} {\bibfnamefont {J.}~\bibnamefont {Abr{\~a}o}},
  \bibinfo {author} {\bibfnamefont {A.}~\bibnamefont {Vieira}}, \bibinfo
  {author} {\bibfnamefont {J.}~\bibnamefont {Mendes}}, \bibinfo {author}
  {\bibfnamefont {R.}~\bibnamefont {Rodr{\'\i}guez-Su{\'a}rez}}, \ and\
  \bibinfo {author} {\bibfnamefont {A.}~\bibnamefont {Azevedo}},\ }\href@noop
  {} {\bibfield  {journal} {\bibinfo  {journal} {Physical Review B}\ }\textbf
  {\bibinfo {volume} {109}},\ \bibinfo {pages} {014420} (\bibinfo {year}
  {2024})}\BibitemShut {NoStop}%
\bibitem [{\citenamefont {Abr{\~a}o}\ \emph {et~al.}(2025)\citenamefont
  {Abr{\~a}o}, \citenamefont {Santos}, \citenamefont {Costa}, \citenamefont
  {Santos}, \citenamefont {Mendes},\ and\ \citenamefont
  {Azevedo}}]{abrao2025anomalous}%
  \BibitemOpen
  \bibfield  {author} {\bibinfo {author} {\bibfnamefont {J.}~\bibnamefont
  {Abr{\~a}o}}, \bibinfo {author} {\bibfnamefont {E.}~\bibnamefont {Santos}},
  \bibinfo {author} {\bibfnamefont {J.}~\bibnamefont {Costa}}, \bibinfo
  {author} {\bibfnamefont {J.}~\bibnamefont {Santos}}, \bibinfo {author}
  {\bibfnamefont {J.}~\bibnamefont {Mendes}}, \ and\ \bibinfo {author}
  {\bibfnamefont {A.}~\bibnamefont {Azevedo}},\ }\href@noop {} {\bibfield
  {journal} {\bibinfo  {journal} {Physical Review Letters}\ }\textbf {\bibinfo
  {volume} {134}},\ \bibinfo {pages} {026702} (\bibinfo {year}
  {2025})}\BibitemShut {NoStop}%
\bibitem [{\citenamefont {Go}\ and\ \citenamefont {Lee}(2020)}]{go2020orbital}%
  \BibitemOpen
  \bibfield  {author} {\bibinfo {author} {\bibfnamefont {D.}~\bibnamefont
  {Go}}\ and\ \bibinfo {author} {\bibfnamefont {H.-W.}\ \bibnamefont {Lee}},\
  }\href@noop {} {\bibfield  {journal} {\bibinfo  {journal} {Physical review
  research}\ }\textbf {\bibinfo {volume} {2}},\ \bibinfo {pages} {013177}
  (\bibinfo {year} {2020})}\BibitemShut {NoStop}%
\bibitem [{\citenamefont {Ding}\ \emph {et~al.}(2020)\citenamefont {Ding},
  \citenamefont {Ross}, \citenamefont {Go}, \citenamefont {Baldrati},
  \citenamefont {Ren}, \citenamefont {Freimuth}, \citenamefont {Becker},
  \citenamefont {Kammerbauer}, \citenamefont {Yang}, \citenamefont {Jakob},
  \citenamefont {Mokrousov},\ and\ \citenamefont {Kl\"aui}}]{Orbital-torque-1}%
  \BibitemOpen
  \bibfield  {author} {\bibinfo {author} {\bibfnamefont {S.}~\bibnamefont
  {Ding}}, \bibinfo {author} {\bibfnamefont {A.}~\bibnamefont {Ross}}, \bibinfo
  {author} {\bibfnamefont {D.}~\bibnamefont {Go}}, \bibinfo {author}
  {\bibfnamefont {L.}~\bibnamefont {Baldrati}}, \bibinfo {author}
  {\bibfnamefont {Z.}~\bibnamefont {Ren}}, \bibinfo {author} {\bibfnamefont
  {F.}~\bibnamefont {Freimuth}}, \bibinfo {author} {\bibfnamefont
  {S.}~\bibnamefont {Becker}}, \bibinfo {author} {\bibfnamefont
  {F.}~\bibnamefont {Kammerbauer}}, \bibinfo {author} {\bibfnamefont
  {J.}~\bibnamefont {Yang}}, \bibinfo {author} {\bibfnamefont {G.}~\bibnamefont
  {Jakob}}, \bibinfo {author} {\bibfnamefont {Y.}~\bibnamefont {Mokrousov}}, \
  and\ \bibinfo {author} {\bibfnamefont {M.}~\bibnamefont {Kl\"aui}},\ }\href
  {\doibase 10.1103/PhysRevLett.125.177201} {\bibfield  {journal} {\bibinfo
  {journal} {Phys. Rev. Lett.}\ }\textbf {\bibinfo {volume} {125}},\ \bibinfo
  {pages} {177201} (\bibinfo {year} {2020})}\BibitemShut {NoStop}%
\bibitem [{\citenamefont {Go}\ \emph {et~al.}(2020)\citenamefont {Go},
  \citenamefont {Freimuth}, \citenamefont {Hanke}, \citenamefont {Xue},
  \citenamefont {Gomonay}, \citenamefont {Lee}, \citenamefont {Bl\"ugel},
  \citenamefont {Haney}, \citenamefont {Lee},\ and\ \citenamefont
  {Mokrousov}}]{Orbital-torque-2}%
  \BibitemOpen
  \bibfield  {author} {\bibinfo {author} {\bibfnamefont {D.}~\bibnamefont
  {Go}}, \bibinfo {author} {\bibfnamefont {F.}~\bibnamefont {Freimuth}},
  \bibinfo {author} {\bibfnamefont {J.-P.}\ \bibnamefont {Hanke}}, \bibinfo
  {author} {\bibfnamefont {F.}~\bibnamefont {Xue}}, \bibinfo {author}
  {\bibfnamefont {O.}~\bibnamefont {Gomonay}}, \bibinfo {author} {\bibfnamefont
  {K.-J.}\ \bibnamefont {Lee}}, \bibinfo {author} {\bibfnamefont
  {S.}~\bibnamefont {Bl\"ugel}}, \bibinfo {author} {\bibfnamefont {P.~M.}\
  \bibnamefont {Haney}}, \bibinfo {author} {\bibfnamefont {H.-W.}\ \bibnamefont
  {Lee}}, \ and\ \bibinfo {author} {\bibfnamefont {Y.}~\bibnamefont
  {Mokrousov}},\ }\href {\doibase 10.1103/PhysRevResearch.2.033401} {\bibfield
  {journal} {\bibinfo  {journal} {Phys. Rev. Research}\ }\textbf {\bibinfo
  {volume} {2}},\ \bibinfo {pages} {033401} (\bibinfo {year}
  {2020})}\BibitemShut {NoStop}%
\bibitem [{\citenamefont {Lee}\ \emph {et~al.}(2021)\citenamefont {Lee},
  \citenamefont {Go}, \citenamefont {Park}, \citenamefont {Jeong},
  \citenamefont {Ko}, \citenamefont {Yun}, \citenamefont {Jo}, \citenamefont
  {Lee}, \citenamefont {Go}, \citenamefont {Oh}, \citenamefont {Kim},
  \citenamefont {Park}, \citenamefont {Min}, \citenamefont {Koo}, \citenamefont
  {Lee}, \citenamefont {Lee},\ and\ \citenamefont
  {Lee}}]{orbital-torque-magnetic-bilayers-EXP}%
  \BibitemOpen
  \bibfield  {author} {\bibinfo {author} {\bibfnamefont {D.}~\bibnamefont
  {Lee}}, \bibinfo {author} {\bibfnamefont {D.}~\bibnamefont {Go}}, \bibinfo
  {author} {\bibfnamefont {H.-J.}\ \bibnamefont {Park}}, \bibinfo {author}
  {\bibfnamefont {W.}~\bibnamefont {Jeong}}, \bibinfo {author} {\bibfnamefont
  {H.-W.}\ \bibnamefont {Ko}}, \bibinfo {author} {\bibfnamefont
  {D.}~\bibnamefont {Yun}}, \bibinfo {author} {\bibfnamefont {D.}~\bibnamefont
  {Jo}}, \bibinfo {author} {\bibfnamefont {S.}~\bibnamefont {Lee}}, \bibinfo
  {author} {\bibfnamefont {G.}~\bibnamefont {Go}}, \bibinfo {author}
  {\bibfnamefont {J.~H.}\ \bibnamefont {Oh}}, \bibinfo {author} {\bibfnamefont
  {K.-J.}\ \bibnamefont {Kim}}, \bibinfo {author} {\bibfnamefont {B.-G.}\
  \bibnamefont {Park}}, \bibinfo {author} {\bibfnamefont {B.-C.}\ \bibnamefont
  {Min}}, \bibinfo {author} {\bibfnamefont {H.~C.}\ \bibnamefont {Koo}},
  \bibinfo {author} {\bibfnamefont {H.-W.}\ \bibnamefont {Lee}}, \bibinfo
  {author} {\bibfnamefont {O.}~\bibnamefont {Lee}}, \ and\ \bibinfo {author}
  {\bibfnamefont {K.-J.}\ \bibnamefont {Lee}},\ }\href {\doibase
  10.1038/s41467-021-26650-9} {\bibfield  {journal} {\bibinfo  {journal}
  {Nature Communications}\ }\textbf {\bibinfo {volume} {12}},\ \bibinfo {pages}
  {6710} (\bibinfo {year} {2021})}\BibitemShut {NoStop}%
\bibitem [{\citenamefont {Fukunaga}\ \emph {et~al.}(2023)\citenamefont
  {Fukunaga}, \citenamefont {Haku}, \citenamefont {Hayashi},\ and\
  \citenamefont {Ando}}]{fukunaga2023orbital}%
  \BibitemOpen
  \bibfield  {author} {\bibinfo {author} {\bibfnamefont {R.}~\bibnamefont
  {Fukunaga}}, \bibinfo {author} {\bibfnamefont {S.}~\bibnamefont {Haku}},
  \bibinfo {author} {\bibfnamefont {H.}~\bibnamefont {Hayashi}}, \ and\
  \bibinfo {author} {\bibfnamefont {K.}~\bibnamefont {Ando}},\ }\href@noop {}
  {\bibfield  {journal} {\bibinfo  {journal} {Physical Review Research}\
  }\textbf {\bibinfo {volume} {5}},\ \bibinfo {pages} {023054} (\bibinfo {year}
  {2023})}\BibitemShut {NoStop}%
\bibitem [{\citenamefont {Gupta}\ \emph {et~al.}(2025)\citenamefont {Gupta},
  \citenamefont {Bouard}, \citenamefont {Kammerbauer}, \citenamefont
  {Ledesma-Martin}, \citenamefont {Bose}, \citenamefont {Kononenko},
  \citenamefont {Martin}, \citenamefont {Us{\'e}}, \citenamefont {Jakob},
  \citenamefont {Drouard} \emph {et~al.}}]{gupta2025harnessing}%
  \BibitemOpen
  \bibfield  {author} {\bibinfo {author} {\bibfnamefont {R.}~\bibnamefont
  {Gupta}}, \bibinfo {author} {\bibfnamefont {C.}~\bibnamefont {Bouard}},
  \bibinfo {author} {\bibfnamefont {F.}~\bibnamefont {Kammerbauer}}, \bibinfo
  {author} {\bibfnamefont {J.~O.}\ \bibnamefont {Ledesma-Martin}}, \bibinfo
  {author} {\bibfnamefont {A.}~\bibnamefont {Bose}}, \bibinfo {author}
  {\bibfnamefont {I.}~\bibnamefont {Kononenko}}, \bibinfo {author}
  {\bibfnamefont {S.}~\bibnamefont {Martin}}, \bibinfo {author} {\bibfnamefont
  {P.}~\bibnamefont {Us{\'e}}}, \bibinfo {author} {\bibfnamefont
  {G.}~\bibnamefont {Jakob}}, \bibinfo {author} {\bibfnamefont
  {M.}~\bibnamefont {Drouard}},  \emph {et~al.},\ }\href@noop {} {\bibfield
  {journal} {\bibinfo  {journal} {Nature Communications}\ }\textbf {\bibinfo
  {volume} {16}},\ \bibinfo {pages} {130} (\bibinfo {year} {2025})}\BibitemShut
  {NoStop}%
\bibitem [{\citenamefont {Matsumoto}\ \emph {et~al.}()\citenamefont
  {Matsumoto}, \citenamefont {Ohshima}, \citenamefont {Ando}, \citenamefont
  {Go}, \citenamefont {Mokrousov},\ and\ \citenamefont
  {Shiraishi}}]{matsumoto2025observation}%
  \BibitemOpen
  \bibfield  {author} {\bibinfo {author} {\bibfnamefont {R.}~\bibnamefont
  {Matsumoto}}, \bibinfo {author} {\bibfnamefont {R.}~\bibnamefont {Ohshima}},
  \bibinfo {author} {\bibfnamefont {Y.}~\bibnamefont {Ando}}, \bibinfo {author}
  {\bibfnamefont {D.}~\bibnamefont {Go}}, \bibinfo {author} {\bibfnamefont
  {Y.}~\bibnamefont {Mokrousov}}, \ and\ \bibinfo {author} {\bibfnamefont
  {M.}~\bibnamefont {Shiraishi}},\ }\href@noop {} {\bibinfo  {journal}
  {arXiv:2501.14237}\ }\BibitemShut {NoStop}%
\bibitem [{\citenamefont {Canonico}\ \emph {et~al.}(2020)\citenamefont
  {Canonico}, \citenamefont {Cysne}, \citenamefont {Molina-Sanchez},
  \citenamefont {Muniz},\ and\ \citenamefont {Rappoport}}]{Us2}%
  \BibitemOpen
\bibfield  {journal} {  }\bibfield  {author} {\bibinfo {author} {\bibfnamefont
  {L.~M.}\ \bibnamefont {Canonico}}, \bibinfo {author} {\bibfnamefont {T.~P.}\
  \bibnamefont {Cysne}}, \bibinfo {author} {\bibfnamefont {A.}~\bibnamefont
  {Molina-Sanchez}}, \bibinfo {author} {\bibfnamefont {R.~B.}\ \bibnamefont
  {Muniz}}, \ and\ \bibinfo {author} {\bibfnamefont {T.~G.}\ \bibnamefont
  {Rappoport}},\ }\href {\doibase 10.1103/PhysRevB.101.161409} {\bibfield
  {journal} {\bibinfo  {journal} {Phys. Rev. B}\ }\textbf {\bibinfo {volume}
  {101}},\ \bibinfo {pages} {161409} (\bibinfo {year} {2020})}\BibitemShut
  {NoStop}%
\bibitem [{\citenamefont {Cysne}\ \emph {et~al.}(2021)\citenamefont {Cysne},
  \citenamefont {Costa}, \citenamefont {Canonico}, \citenamefont {Nardelli},
  \citenamefont {Muniz},\ and\ \citenamefont {Rappoport}}]{Us3}%
  \BibitemOpen
  \bibfield  {author} {\bibinfo {author} {\bibfnamefont {T.~P.}\ \bibnamefont
  {Cysne}}, \bibinfo {author} {\bibfnamefont {M.}~\bibnamefont {Costa}},
  \bibinfo {author} {\bibfnamefont {L.~M.}\ \bibnamefont {Canonico}}, \bibinfo
  {author} {\bibfnamefont {M.~B.}\ \bibnamefont {Nardelli}}, \bibinfo {author}
  {\bibfnamefont {R.~B.}\ \bibnamefont {Muniz}}, \ and\ \bibinfo {author}
  {\bibfnamefont {T.~G.}\ \bibnamefont {Rappoport}},\ }\href {\doibase
  10.1103/PhysRevLett.126.056601} {\bibfield  {journal} {\bibinfo  {journal}
  {Phys. Rev. Lett.}\ }\textbf {\bibinfo {volume} {126}},\ \bibinfo {pages}
  {056601} (\bibinfo {year} {2021})}\BibitemShut {NoStop}%
\bibitem [{\citenamefont {Bhowal}\ and\ \citenamefont
  {Satpathy}(2020)}]{OHE_Bhowal_1}%
  \BibitemOpen
  \bibfield  {author} {\bibinfo {author} {\bibfnamefont {S.}~\bibnamefont
  {Bhowal}}\ and\ \bibinfo {author} {\bibfnamefont {S.}~\bibnamefont
  {Satpathy}},\ }\href {\doibase 10.1103/PhysRevB.102.035409} {\bibfield
  {journal} {\bibinfo  {journal} {Phys. Rev. B}\ }\textbf {\bibinfo {volume}
  {102}},\ \bibinfo {pages} {035409} (\bibinfo {year} {2020})}\BibitemShut
  {NoStop}%
\bibitem [{\citenamefont {Costa}\ \emph {et~al.}(2023)\citenamefont {Costa},
  \citenamefont {Focassio}, \citenamefont {Canonico}, \citenamefont {Cysne},
  \citenamefont {Schleder}, \citenamefont {Muniz}, \citenamefont {Fazzio},\
  and\ \citenamefont {Rappoport}}]{costa2022connecting}%
  \BibitemOpen
  \bibfield  {author} {\bibinfo {author} {\bibfnamefont {M.}~\bibnamefont
  {Costa}}, \bibinfo {author} {\bibfnamefont {B.}~\bibnamefont {Focassio}},
  \bibinfo {author} {\bibfnamefont {L.~M.}\ \bibnamefont {Canonico}}, \bibinfo
  {author} {\bibfnamefont {T.~P.}\ \bibnamefont {Cysne}}, \bibinfo {author}
  {\bibfnamefont {G.~R.}\ \bibnamefont {Schleder}}, \bibinfo {author}
  {\bibfnamefont {R.~B.}\ \bibnamefont {Muniz}}, \bibinfo {author}
  {\bibfnamefont {A.}~\bibnamefont {Fazzio}}, \ and\ \bibinfo {author}
  {\bibfnamefont {T.~G.}\ \bibnamefont {Rappoport}},\ }\href {\doibase
  10.1103/PhysRevLett.130.116204} {\bibfield  {journal} {\bibinfo  {journal}
  {Phys. Rev. Lett.}\ }\textbf {\bibinfo {volume} {130}},\ \bibinfo {pages}
  {116204} (\bibinfo {year} {2023})}\BibitemShut {NoStop}%
\bibitem [{\citenamefont {Cysne}\ \emph {et~al.}()\citenamefont {Cysne},
  \citenamefont {Canonico}, \citenamefont {Costa}, \citenamefont {Muniz},\ and\
  \citenamefont {Rappoport}}]{cysne2025orbitronicsreview}%
  \BibitemOpen
  \bibfield  {author} {\bibinfo {author} {\bibfnamefont {T.~P.}\ \bibnamefont
  {Cysne}}, \bibinfo {author} {\bibfnamefont {L.~M.}\ \bibnamefont {Canonico}},
  \bibinfo {author} {\bibfnamefont {M.}~\bibnamefont {Costa}}, \bibinfo
  {author} {\bibfnamefont {R.}~\bibnamefont {Muniz}}, \ and\ \bibinfo {author}
  {\bibfnamefont {T.~G.}\ \bibnamefont {Rappoport}},\ }\href@noop {} {\bibinfo
  {journal} {arXiv:2502.12339}\ }\BibitemShut {NoStop}%
\bibitem [{\citenamefont {Canonico}\ \emph {et~al.}()\citenamefont {Canonico},
  \citenamefont {Garc{\'\i}a},\ and\ \citenamefont {Roche}}]{canonico2023spin}%
  \BibitemOpen
\bibfield  {journal} {  }\bibfield  {author} {\bibinfo {author} {\bibfnamefont
  {L.~M.}\ \bibnamefont {Canonico}}, \bibinfo {author} {\bibfnamefont {J.~H.}\
  \bibnamefont {Garc{\'\i}a}}, \ and\ \bibinfo {author} {\bibfnamefont
  {S.}~\bibnamefont {Roche}},\ }\href@noop {} {\bibinfo  {journal}
  {arXiv:2307.14673}\ }\BibitemShut {NoStop}%
\bibitem [{\citenamefont {Hanke}\ \emph {et~al.}(2016)\citenamefont {Hanke},
  \citenamefont {Freimuth}, \citenamefont {Nandy}, \citenamefont {Zhang},
  \citenamefont {Bl\"ugel},\ and\ \citenamefont
  {Mokrousov}}]{PRoleBerryMoment}%
  \BibitemOpen
\bibfield  {journal} {  }\bibfield  {author} {\bibinfo {author} {\bibfnamefont
  {J.-P.}\ \bibnamefont {Hanke}}, \bibinfo {author} {\bibfnamefont
  {F.}~\bibnamefont {Freimuth}}, \bibinfo {author} {\bibfnamefont {A.~K.}\
  \bibnamefont {Nandy}}, \bibinfo {author} {\bibfnamefont {H.}~\bibnamefont
  {Zhang}}, \bibinfo {author} {\bibfnamefont {S.}~\bibnamefont {Bl\"ugel}}, \
  and\ \bibinfo {author} {\bibfnamefont {Y.}~\bibnamefont {Mokrousov}},\ }\href
  {\doibase 10.1103/PhysRevB.94.121114} {\bibfield  {journal} {\bibinfo
  {journal} {Phys. Rev. B}\ }\textbf {\bibinfo {volume} {94}},\ \bibinfo
  {pages} {121114} (\bibinfo {year} {2016})}\BibitemShut {NoStop}%
\bibitem [{\citenamefont {Bhowal}\ and\ \citenamefont
  {Vignale}(2021)}]{OHE_Bhowal-Vignale}%
  \BibitemOpen
  \bibfield  {author} {\bibinfo {author} {\bibfnamefont {S.}~\bibnamefont
  {Bhowal}}\ and\ \bibinfo {author} {\bibfnamefont {G.}~\bibnamefont
  {Vignale}},\ }\href {\doibase 10.1103/PhysRevB.103.195309} {\bibfield
  {journal} {\bibinfo  {journal} {Phys. Rev. B}\ }\textbf {\bibinfo {volume}
  {103}},\ \bibinfo {pages} {195309} (\bibinfo {year} {2021})}\BibitemShut
  {NoStop}%
\bibitem [{\citenamefont {Salvador-S\'anchez}\ \emph
  {et~al.}(2024)\citenamefont {Salvador-S\'anchez}, \citenamefont {Canonico},
  \citenamefont {P\'erez-Rodr\'{\i}guez}, \citenamefont {Cysne}, \citenamefont
  {Baba}, \citenamefont {Cleric\`o}, \citenamefont {Vila}, \citenamefont
  {Vaquero}, \citenamefont {Delgado-Notario}, \citenamefont {Caridad},
  \citenamefont {Watanabe}, \citenamefont {Taniguchi}, \citenamefont {Molina},
  \citenamefont {Dom\'{\i}nguez-Adame}, \citenamefont {Roche}, \citenamefont
  {Diez}, \citenamefont {Rappoport},\ and\ \citenamefont
  {Amado}}]{nonlocalMario}%
  \BibitemOpen
  \bibfield  {author} {\bibinfo {author} {\bibfnamefont {J.}~\bibnamefont
  {Salvador-S\'anchez}}, \bibinfo {author} {\bibfnamefont {L.~M.}\ \bibnamefont
  {Canonico}}, \bibinfo {author} {\bibfnamefont {A.}~\bibnamefont
  {P\'erez-Rodr\'{\i}guez}}, \bibinfo {author} {\bibfnamefont {T.~P.}\
  \bibnamefont {Cysne}}, \bibinfo {author} {\bibfnamefont {Y.}~\bibnamefont
  {Baba}}, \bibinfo {author} {\bibfnamefont {V.}~\bibnamefont {Cleric\`o}},
  \bibinfo {author} {\bibfnamefont {M.}~\bibnamefont {Vila}}, \bibinfo {author}
  {\bibfnamefont {D.}~\bibnamefont {Vaquero}}, \bibinfo {author} {\bibfnamefont
  {J.~A.}\ \bibnamefont {Delgado-Notario}}, \bibinfo {author} {\bibfnamefont
  {J.~M.}\ \bibnamefont {Caridad}}, \bibinfo {author} {\bibfnamefont
  {K.}~\bibnamefont {Watanabe}}, \bibinfo {author} {\bibfnamefont
  {T.}~\bibnamefont {Taniguchi}}, \bibinfo {author} {\bibfnamefont {R.~A.}\
  \bibnamefont {Molina}}, \bibinfo {author} {\bibfnamefont {F.}~\bibnamefont
  {Dom\'{\i}nguez-Adame}}, \bibinfo {author} {\bibfnamefont {S.}~\bibnamefont
  {Roche}}, \bibinfo {author} {\bibfnamefont {E.}~\bibnamefont {Diez}},
  \bibinfo {author} {\bibfnamefont {T.~G.}\ \bibnamefont {Rappoport}}, \ and\
  \bibinfo {author} {\bibfnamefont {M.}~\bibnamefont {Amado}},\ }\href
  {\doibase 10.1103/PhysRevResearch.6.023212} {\bibfield  {journal} {\bibinfo
  {journal} {Phys. Rev. Res.}\ }\textbf {\bibinfo {volume} {6}},\ \bibinfo
  {pages} {023212} (\bibinfo {year} {2024})}\BibitemShut {NoStop}%
\bibitem [{\citenamefont {Schaefer}\ and\ \citenamefont
  {Nowack}(2021)}]{TunableReversableMagnetoelectricBG}%
  \BibitemOpen
  \bibfield  {author} {\bibinfo {author} {\bibfnamefont {B.~T.}\ \bibnamefont
  {Schaefer}}\ and\ \bibinfo {author} {\bibfnamefont {K.~C.}\ \bibnamefont
  {Nowack}},\ }\href {\doibase 10.1103/PhysRevB.103.224426} {\bibfield
  {journal} {\bibinfo  {journal} {Phys. Rev. B}\ }\textbf {\bibinfo {volume}
  {103}},\ \bibinfo {pages} {224426} (\bibinfo {year} {2021})}\BibitemShut
  {NoStop}%
\bibitem [{\citenamefont {Cysne}\ \emph {et~al.}(2024)\citenamefont {Cysne},
  \citenamefont {Kort-Kamp},\ and\ \citenamefont
  {Rappoport}}]{cysne2024controlling}%
  \BibitemOpen
  \bibfield  {author} {\bibinfo {author} {\bibfnamefont {T.~P.}\ \bibnamefont
  {Cysne}}, \bibinfo {author} {\bibfnamefont {W.~J.~M.}\ \bibnamefont
  {Kort-Kamp}}, \ and\ \bibinfo {author} {\bibfnamefont {T.~G.}\ \bibnamefont
  {Rappoport}},\ }\href {\doibase 10.1103/PhysRevResearch.6.023271} {\bibfield
  {journal} {\bibinfo  {journal} {Phys. Rev. Res.}\ }\textbf {\bibinfo {volume}
  {6}},\ \bibinfo {pages} {023271} (\bibinfo {year} {2024})}\BibitemShut
  {NoStop}%
\bibitem [{\citenamefont {He}\ \emph {et~al.}(2020)\citenamefont {He},
  \citenamefont {Goldhaber-Gordon},\ and\ \citenamefont {Law}}]{he2020giant}%
  \BibitemOpen
  \bibfield  {author} {\bibinfo {author} {\bibfnamefont {W.-Y.}\ \bibnamefont
  {He}}, \bibinfo {author} {\bibfnamefont {D.}~\bibnamefont
  {Goldhaber-Gordon}}, \ and\ \bibinfo {author} {\bibfnamefont {K.~T.}\
  \bibnamefont {Law}},\ }\href {\doibase 10.1038/s41467-020-15473-9} {\bibfield
   {journal} {\bibinfo  {journal} {Nature Communications}\ }\textbf {\bibinfo
  {volume} {11}},\ \bibinfo {pages} {1650} (\bibinfo {year}
  {2020})}\BibitemShut {NoStop}%
\bibitem [{\citenamefont {Serlin}\ \emph {et~al.}(2020)\citenamefont {Serlin},
  \citenamefont {Tschirhart}, \citenamefont {Polshyn}, \citenamefont {Zhang},
  \citenamefont {Zhu}, \citenamefont {Watanabe}, \citenamefont {Taniguchi},
  \citenamefont {Balents},\ and\ \citenamefont {Young}}]{serlin2020intrinsic}%
  \BibitemOpen
  \bibfield  {author} {\bibinfo {author} {\bibfnamefont {M.}~\bibnamefont
  {Serlin}}, \bibinfo {author} {\bibfnamefont {C.}~\bibnamefont {Tschirhart}},
  \bibinfo {author} {\bibfnamefont {H.}~\bibnamefont {Polshyn}}, \bibinfo
  {author} {\bibfnamefont {Y.}~\bibnamefont {Zhang}}, \bibinfo {author}
  {\bibfnamefont {J.}~\bibnamefont {Zhu}}, \bibinfo {author} {\bibfnamefont
  {K.}~\bibnamefont {Watanabe}}, \bibinfo {author} {\bibfnamefont
  {T.}~\bibnamefont {Taniguchi}}, \bibinfo {author} {\bibfnamefont
  {L.}~\bibnamefont {Balents}}, \ and\ \bibinfo {author} {\bibfnamefont
  {A.}~\bibnamefont {Young}},\ }\href {\doibase 10.1126/science.aay5533}
  {\bibfield  {journal} {\bibinfo  {journal} {Science}\ }\textbf {\bibinfo
  {volume} {367}},\ \bibinfo {pages} {900} (\bibinfo {year}
  {2020})}\BibitemShut {NoStop}%
\bibitem [{\citenamefont {Sodemann}\ and\ \citenamefont
  {Fu}(2015)}]{sodemann2015quantum}%
  \BibitemOpen
  \bibfield  {author} {\bibinfo {author} {\bibfnamefont {I.}~\bibnamefont
  {Sodemann}}\ and\ \bibinfo {author} {\bibfnamefont {L.}~\bibnamefont {Fu}},\
  }\href@noop {} {\bibfield  {journal} {\bibinfo  {journal} {Physical review
  letters}\ }\textbf {\bibinfo {volume} {115}},\ \bibinfo {pages} {216806}
  (\bibinfo {year} {2015})}\BibitemShut {NoStop}%
\bibitem [{\citenamefont {Shi}\ and\ \citenamefont
  {Song}(2019)}]{shi2019symmetry}%
  \BibitemOpen
  \bibfield  {author} {\bibinfo {author} {\bibfnamefont {L.-k.}\ \bibnamefont
  {Shi}}\ and\ \bibinfo {author} {\bibfnamefont {J.~C.}\ \bibnamefont {Song}},\
  }\href@noop {} {\bibfield  {journal} {\bibinfo  {journal} {Physical Review
  B}\ }\textbf {\bibinfo {volume} {99}},\ \bibinfo {pages} {035403} (\bibinfo
  {year} {2019})}\BibitemShut {NoStop}%
\bibitem [{\citenamefont {Ye}\ \emph {et~al.}(2024)\citenamefont {Ye},
  \citenamefont {Zhu}, \citenamefont {Xu}, \citenamefont {Zhao},\ and\
  \citenamefont {Liao}}]{ye2024nonlinear}%
  \BibitemOpen
  \bibfield  {author} {\bibinfo {author} {\bibfnamefont {X.-G.}\ \bibnamefont
  {Ye}}, \bibinfo {author} {\bibfnamefont {P.-F.}\ \bibnamefont {Zhu}},
  \bibinfo {author} {\bibfnamefont {W.-Z.}\ \bibnamefont {Xu}}, \bibinfo
  {author} {\bibfnamefont {T.-Y.}\ \bibnamefont {Zhao}}, \ and\ \bibinfo
  {author} {\bibfnamefont {Z.-M.}\ \bibnamefont {Liao}},\ }\href@noop {}
  {\bibfield  {journal} {\bibinfo  {journal} {Physical Review B}\ }\textbf
  {\bibinfo {volume} {110}},\ \bibinfo {pages} {L201407} (\bibinfo {year}
  {2024})}\BibitemShut {NoStop}%
\bibitem [{\citenamefont {Ovalle}\ \emph {et~al.}(2024)\citenamefont {Ovalle},
  \citenamefont {Pezo},\ and\ \citenamefont {Manchon}}]{ovalle2024orbital}%
  \BibitemOpen
  \bibfield  {author} {\bibinfo {author} {\bibfnamefont {D.~G.}\ \bibnamefont
  {Ovalle}}, \bibinfo {author} {\bibfnamefont {A.}~\bibnamefont {Pezo}}, \ and\
  \bibinfo {author} {\bibfnamefont {A.}~\bibnamefont {Manchon}},\ }\href@noop
  {} {\bibfield  {journal} {\bibinfo  {journal} {Physical Review B}\ }\textbf
  {\bibinfo {volume} {110}},\ \bibinfo {pages} {094439} (\bibinfo {year}
  {2024})}\BibitemShut {NoStop}%
\bibitem [{\citenamefont {Li}\ \emph {et~al.}(2024)\citenamefont {Li},
  \citenamefont {Liu}, \citenamefont {Pan}, \citenamefont {Wang}, \citenamefont
  {Zhang}, \citenamefont {Yu},\ and\ \citenamefont {Liao}}]{li2024room}%
  \BibitemOpen
  \bibfield  {author} {\bibinfo {author} {\bibfnamefont {D.}~\bibnamefont
  {Li}}, \bibinfo {author} {\bibfnamefont {X.-Y.}\ \bibnamefont {Liu}},
  \bibinfo {author} {\bibfnamefont {Z.-C.}\ \bibnamefont {Pan}}, \bibinfo
  {author} {\bibfnamefont {A.-Q.}\ \bibnamefont {Wang}}, \bibinfo {author}
  {\bibfnamefont {J.}~\bibnamefont {Zhang}}, \bibinfo {author} {\bibfnamefont
  {P.}~\bibnamefont {Yu}}, \ and\ \bibinfo {author} {\bibfnamefont {Z.-M.}\
  \bibnamefont {Liao}},\ }\href@noop {} {\bibfield  {journal} {\bibinfo
  {journal} {Physical Review B}\ }\textbf {\bibinfo {volume} {110}},\ \bibinfo
  {pages} {035423} (\bibinfo {year} {2024})}\BibitemShut {NoStop}%
\bibitem [{\citenamefont {Wang}\ \emph
  {et~al.}(2024{\natexlab{a}})\citenamefont {Wang}, \citenamefont {Li},
  \citenamefont {Zhao}, \citenamefont {Liu}, \citenamefont {Zhang},
  \citenamefont {Liao}, \citenamefont {Yin}, \citenamefont {Pan}, \citenamefont
  {Yu},\ and\ \citenamefont {Liao}}]{wang2024orbital}%
  \BibitemOpen
  \bibfield  {author} {\bibinfo {author} {\bibfnamefont {A.-Q.}\ \bibnamefont
  {Wang}}, \bibinfo {author} {\bibfnamefont {D.}~\bibnamefont {Li}}, \bibinfo
  {author} {\bibfnamefont {T.-Y.}\ \bibnamefont {Zhao}}, \bibinfo {author}
  {\bibfnamefont {X.-Y.}\ \bibnamefont {Liu}}, \bibinfo {author} {\bibfnamefont
  {J.}~\bibnamefont {Zhang}}, \bibinfo {author} {\bibfnamefont
  {X.}~\bibnamefont {Liao}}, \bibinfo {author} {\bibfnamefont {Q.}~\bibnamefont
  {Yin}}, \bibinfo {author} {\bibfnamefont {Z.-C.}\ \bibnamefont {Pan}},
  \bibinfo {author} {\bibfnamefont {P.}~\bibnamefont {Yu}}, \ and\ \bibinfo
  {author} {\bibfnamefont {Z.-M.}\ \bibnamefont {Liao}},\ }\href@noop {}
  {\bibfield  {journal} {\bibinfo  {journal} {Physical Review B}\ }\textbf
  {\bibinfo {volume} {110}},\ \bibinfo {pages} {155434} (\bibinfo {year}
  {2024}{\natexlab{a}})}\BibitemShut {NoStop}%
\bibitem [{\citenamefont {MacNeill}\ \emph {et~al.}(2017)\citenamefont
  {MacNeill}, \citenamefont {Stiehl}, \citenamefont {Guimaraes}, \citenamefont
  {Buhrman}, \citenamefont {Park},\ and\ \citenamefont
  {Ralph}}]{macneill2017control}%
  \BibitemOpen
  \bibfield  {author} {\bibinfo {author} {\bibfnamefont {D.}~\bibnamefont
  {MacNeill}}, \bibinfo {author} {\bibfnamefont {G.}~\bibnamefont {Stiehl}},
  \bibinfo {author} {\bibfnamefont {M.}~\bibnamefont {Guimaraes}}, \bibinfo
  {author} {\bibfnamefont {R.}~\bibnamefont {Buhrman}}, \bibinfo {author}
  {\bibfnamefont {J.}~\bibnamefont {Park}}, \ and\ \bibinfo {author}
  {\bibfnamefont {D.}~\bibnamefont {Ralph}},\ }\href@noop {} {\bibfield
  {journal} {\bibinfo  {journal} {Nature Physics}\ }\textbf {\bibinfo {volume}
  {13}},\ \bibinfo {pages} {300} (\bibinfo {year} {2017})}\BibitemShut
  {NoStop}%
\bibitem [{\citenamefont {Kao}\ \emph {et~al.}(2022)\citenamefont {Kao},
  \citenamefont {Muzzio}, \citenamefont {Zhang}, \citenamefont {Zhu},
  \citenamefont {Gobbo}, \citenamefont {Yuan}, \citenamefont {Weber},
  \citenamefont {Rao}, \citenamefont {Li}, \citenamefont {Edgar} \emph
  {et~al.}}]{kao2022deterministic}%
  \BibitemOpen
  \bibfield  {author} {\bibinfo {author} {\bibfnamefont {I.-H.}\ \bibnamefont
  {Kao}}, \bibinfo {author} {\bibfnamefont {R.}~\bibnamefont {Muzzio}},
  \bibinfo {author} {\bibfnamefont {H.}~\bibnamefont {Zhang}}, \bibinfo
  {author} {\bibfnamefont {M.}~\bibnamefont {Zhu}}, \bibinfo {author}
  {\bibfnamefont {J.}~\bibnamefont {Gobbo}}, \bibinfo {author} {\bibfnamefont
  {S.}~\bibnamefont {Yuan}}, \bibinfo {author} {\bibfnamefont {D.}~\bibnamefont
  {Weber}}, \bibinfo {author} {\bibfnamefont {R.}~\bibnamefont {Rao}}, \bibinfo
  {author} {\bibfnamefont {J.}~\bibnamefont {Li}}, \bibinfo {author}
  {\bibfnamefont {J.~H.}\ \bibnamefont {Edgar}},  \emph {et~al.},\ }\href@noop
  {} {\bibfield  {journal} {\bibinfo  {journal} {Nature materials}\ }\textbf
  {\bibinfo {volume} {21}},\ \bibinfo {pages} {1029} (\bibinfo {year}
  {2022})}\BibitemShut {NoStop}%
\bibitem [{\citenamefont {Zhang}\ \emph {et~al.}(2023)\citenamefont {Zhang},
  \citenamefont {Xu}, \citenamefont {Jia}, \citenamefont {Lan}, \citenamefont
  {Huang}, \citenamefont {He}, \citenamefont {He}, \citenamefont {Shao},
  \citenamefont {Wang}, \citenamefont {Zhao} \emph {et~al.}}]{zhang2023room}%
  \BibitemOpen
  \bibfield  {author} {\bibinfo {author} {\bibfnamefont {Y.}~\bibnamefont
  {Zhang}}, \bibinfo {author} {\bibfnamefont {H.}~\bibnamefont {Xu}}, \bibinfo
  {author} {\bibfnamefont {K.}~\bibnamefont {Jia}}, \bibinfo {author}
  {\bibfnamefont {G.}~\bibnamefont {Lan}}, \bibinfo {author} {\bibfnamefont
  {Z.}~\bibnamefont {Huang}}, \bibinfo {author} {\bibfnamefont
  {B.}~\bibnamefont {He}}, \bibinfo {author} {\bibfnamefont {C.}~\bibnamefont
  {He}}, \bibinfo {author} {\bibfnamefont {Q.}~\bibnamefont {Shao}}, \bibinfo
  {author} {\bibfnamefont {Y.}~\bibnamefont {Wang}}, \bibinfo {author}
  {\bibfnamefont {M.}~\bibnamefont {Zhao}},  \emph {et~al.},\ }\href@noop {}
  {\bibfield  {journal} {\bibinfo  {journal} {Science Advances}\ }\textbf
  {\bibinfo {volume} {9}},\ \bibinfo {pages} {eadg9819} (\bibinfo {year}
  {2023})}\BibitemShut {NoStop}%
\bibitem [{\citenamefont {Liu}\ \emph {et~al.}(2023)\citenamefont {Liu},
  \citenamefont {Shi}, \citenamefont {Kumar}, \citenamefont {Kim},
  \citenamefont {Shi}, \citenamefont {Yang}, \citenamefont {Zhang},
  \citenamefont {Zhang}, \citenamefont {Wang}, \citenamefont {Yang} \emph
  {et~al.}}]{liu2023field}%
  \BibitemOpen
  \bibfield  {author} {\bibinfo {author} {\bibfnamefont {Y.}~\bibnamefont
  {Liu}}, \bibinfo {author} {\bibfnamefont {G.}~\bibnamefont {Shi}}, \bibinfo
  {author} {\bibfnamefont {D.}~\bibnamefont {Kumar}}, \bibinfo {author}
  {\bibfnamefont {T.}~\bibnamefont {Kim}}, \bibinfo {author} {\bibfnamefont
  {S.}~\bibnamefont {Shi}}, \bibinfo {author} {\bibfnamefont {D.}~\bibnamefont
  {Yang}}, \bibinfo {author} {\bibfnamefont {J.}~\bibnamefont {Zhang}},
  \bibinfo {author} {\bibfnamefont {C.}~\bibnamefont {Zhang}}, \bibinfo
  {author} {\bibfnamefont {F.}~\bibnamefont {Wang}}, \bibinfo {author}
  {\bibfnamefont {S.}~\bibnamefont {Yang}},  \emph {et~al.},\ }\href@noop {}
  {\bibfield  {journal} {\bibinfo  {journal} {Nature Electronics}\ }\textbf
  {\bibinfo {volume} {6}},\ \bibinfo {pages} {732} (\bibinfo {year}
  {2023})}\BibitemShut {NoStop}%
\bibitem [{\citenamefont {Wei}\ \emph {et~al.}(2023)\citenamefont {Wei},
  \citenamefont {Yin}, \citenamefont {Liu}, \citenamefont {Zhang},
  \citenamefont {Niu}, \citenamefont {Liu}, \citenamefont {Yang}, \citenamefont
  {Peng}, \citenamefont {Huang}, \citenamefont {Liu} \emph
  {et~al.}}]{wei2023field}%
  \BibitemOpen
  \bibfield  {author} {\bibinfo {author} {\bibfnamefont {L.}~\bibnamefont
  {Wei}}, \bibinfo {author} {\bibfnamefont {X.}~\bibnamefont {Yin}}, \bibinfo
  {author} {\bibfnamefont {P.}~\bibnamefont {Liu}}, \bibinfo {author}
  {\bibfnamefont {P.}~\bibnamefont {Zhang}}, \bibinfo {author} {\bibfnamefont
  {W.}~\bibnamefont {Niu}}, \bibinfo {author} {\bibfnamefont {P.}~\bibnamefont
  {Liu}}, \bibinfo {author} {\bibfnamefont {J.}~\bibnamefont {Yang}}, \bibinfo
  {author} {\bibfnamefont {J.}~\bibnamefont {Peng}}, \bibinfo {author}
  {\bibfnamefont {F.}~\bibnamefont {Huang}}, \bibinfo {author} {\bibfnamefont
  {R.}~\bibnamefont {Liu}},  \emph {et~al.},\ }\href@noop {} {\bibfield
  {journal} {\bibinfo  {journal} {Applied Physics Letters}\ }\textbf {\bibinfo
  {volume} {123}} (\bibinfo {year} {2023})}\BibitemShut {NoStop}%
\bibitem [{\citenamefont {Kajale}\ \emph {et~al.}(2024)\citenamefont {Kajale},
  \citenamefont {Nguyen}, \citenamefont {Hung}, \citenamefont {Li},\ and\
  \citenamefont {Sarkar}}]{kajale2024field}%
  \BibitemOpen
  \bibfield  {author} {\bibinfo {author} {\bibfnamefont {S.~N.}\ \bibnamefont
  {Kajale}}, \bibinfo {author} {\bibfnamefont {T.}~\bibnamefont {Nguyen}},
  \bibinfo {author} {\bibfnamefont {N.~T.}\ \bibnamefont {Hung}}, \bibinfo
  {author} {\bibfnamefont {M.}~\bibnamefont {Li}}, \ and\ \bibinfo {author}
  {\bibfnamefont {D.}~\bibnamefont {Sarkar}},\ }\href@noop {} {\bibfield
  {journal} {\bibinfo  {journal} {Science advances}\ }\textbf {\bibinfo
  {volume} {10}},\ \bibinfo {pages} {eadk8669} (\bibinfo {year}
  {2024})}\BibitemShut {NoStop}%
\bibitem [{\citenamefont {Pu}\ \emph {et~al.}(2024)\citenamefont {Pu},
  \citenamefont {Shi}, \citenamefont {Yang}, \citenamefont {Yang},
  \citenamefont {Wang}, \citenamefont {Zhang},\ and\ \citenamefont
  {Yang}}]{pu2024field}%
  \BibitemOpen
  \bibfield  {author} {\bibinfo {author} {\bibfnamefont {Y.}~\bibnamefont
  {Pu}}, \bibinfo {author} {\bibfnamefont {G.}~\bibnamefont {Shi}}, \bibinfo
  {author} {\bibfnamefont {Q.}~\bibnamefont {Yang}}, \bibinfo {author}
  {\bibfnamefont {D.}~\bibnamefont {Yang}}, \bibinfo {author} {\bibfnamefont
  {F.}~\bibnamefont {Wang}}, \bibinfo {author} {\bibfnamefont {C.}~\bibnamefont
  {Zhang}}, \ and\ \bibinfo {author} {\bibfnamefont {H.}~\bibnamefont {Yang}},\
  }\href@noop {} {\bibfield  {journal} {\bibinfo  {journal} {Advanced
  Functional Materials}\ }\textbf {\bibinfo {volume} {34}},\ \bibinfo {pages}
  {2400143} (\bibinfo {year} {2024})}\BibitemShut {NoStop}%
\bibitem [{\citenamefont {Wang}\ \emph
  {et~al.}(2024{\natexlab{b}})\citenamefont {Wang}, \citenamefont {Shi},
  \citenamefont {Kim}, \citenamefont {Park}, \citenamefont {Jang},
  \citenamefont {Tan}, \citenamefont {Lin}, \citenamefont {Liu}, \citenamefont
  {Kim}, \citenamefont {Yang} \emph {et~al.}}]{wang2024field}%
  \BibitemOpen
  \bibfield  {author} {\bibinfo {author} {\bibfnamefont {F.}~\bibnamefont
  {Wang}}, \bibinfo {author} {\bibfnamefont {G.}~\bibnamefont {Shi}}, \bibinfo
  {author} {\bibfnamefont {K.-W.}\ \bibnamefont {Kim}}, \bibinfo {author}
  {\bibfnamefont {H.-J.}\ \bibnamefont {Park}}, \bibinfo {author}
  {\bibfnamefont {J.~G.}\ \bibnamefont {Jang}}, \bibinfo {author}
  {\bibfnamefont {H.~R.}\ \bibnamefont {Tan}}, \bibinfo {author} {\bibfnamefont
  {M.}~\bibnamefont {Lin}}, \bibinfo {author} {\bibfnamefont {Y.}~\bibnamefont
  {Liu}}, \bibinfo {author} {\bibfnamefont {T.}~\bibnamefont {Kim}}, \bibinfo
  {author} {\bibfnamefont {D.}~\bibnamefont {Yang}},  \emph {et~al.},\
  }\href@noop {} {\bibfield  {journal} {\bibinfo  {journal} {Nature materials}\
  }\textbf {\bibinfo {volume} {23}},\ \bibinfo {pages} {768} (\bibinfo {year}
  {2024}{\natexlab{b}})}\BibitemShut {NoStop}%
\bibitem [{\citenamefont {Pandey}\ \emph {et~al.}()\citenamefont {Pandey},
  \citenamefont {Zhao}, \citenamefont {Tenzin}, \citenamefont {Ngaloy},
  \citenamefont {Lamparsk{\'a}}, \citenamefont {Bangar}, \citenamefont {Ali},
  \citenamefont {Abdel-Hafiez}, \citenamefont {Zhang}, \citenamefont {Wu} \emph
  {et~al.}}]{pandey2024energy}%
  \BibitemOpen
  \bibfield  {author} {\bibinfo {author} {\bibfnamefont {L.}~\bibnamefont
  {Pandey}}, \bibinfo {author} {\bibfnamefont {B.}~\bibnamefont {Zhao}},
  \bibinfo {author} {\bibfnamefont {K.}~\bibnamefont {Tenzin}}, \bibinfo
  {author} {\bibfnamefont {R.}~\bibnamefont {Ngaloy}}, \bibinfo {author}
  {\bibfnamefont {V.}~\bibnamefont {Lamparsk{\'a}}}, \bibinfo {author}
  {\bibfnamefont {H.}~\bibnamefont {Bangar}}, \bibinfo {author} {\bibfnamefont
  {A.}~\bibnamefont {Ali}}, \bibinfo {author} {\bibfnamefont {M.}~\bibnamefont
  {Abdel-Hafiez}}, \bibinfo {author} {\bibfnamefont {G.}~\bibnamefont {Zhang}},
  \bibinfo {author} {\bibfnamefont {H.}~\bibnamefont {Wu}},  \emph {et~al.},\
  }\href@noop {} {\bibinfo  {journal} {arXiv:2408.13095}\ }\BibitemShut
  {NoStop}%
\bibitem [{\citenamefont {Pan}\ \emph {et~al.}(2023)\citenamefont {Pan},
  \citenamefont {Li}, \citenamefont {Ye}, \citenamefont {Chen}, \citenamefont
  {Chen}, \citenamefont {Wang}, \citenamefont {Tian}, \citenamefont {Yao},
  \citenamefont {Liu},\ and\ \citenamefont {Liao}}]{pan2023room}%
  \BibitemOpen
\bibfield  {journal} {  }\bibfield  {author} {\bibinfo {author} {\bibfnamefont
  {Z.-C.}\ \bibnamefont {Pan}}, \bibinfo {author} {\bibfnamefont
  {D.}~\bibnamefont {Li}}, \bibinfo {author} {\bibfnamefont {X.-G.}\
  \bibnamefont {Ye}}, \bibinfo {author} {\bibfnamefont {Z.}~\bibnamefont
  {Chen}}, \bibinfo {author} {\bibfnamefont {Z.-H.}\ \bibnamefont {Chen}},
  \bibinfo {author} {\bibfnamefont {A.-Q.}\ \bibnamefont {Wang}}, \bibinfo
  {author} {\bibfnamefont {M.}~\bibnamefont {Tian}}, \bibinfo {author}
  {\bibfnamefont {G.}~\bibnamefont {Yao}}, \bibinfo {author} {\bibfnamefont
  {K.}~\bibnamefont {Liu}}, \ and\ \bibinfo {author} {\bibfnamefont {Z.-M.}\
  \bibnamefont {Liao}},\ }\href@noop {} {\bibfield  {journal} {\bibinfo
  {journal} {Science Bulletin}\ }\textbf {\bibinfo {volume} {68}},\ \bibinfo
  {pages} {2743} (\bibinfo {year} {2023})}\BibitemShut {NoStop}%
\bibitem [{\citenamefont {Canonico}\ \emph {et~al.}(2024)\citenamefont
  {Canonico}, \citenamefont {Garcia},\ and\ \citenamefont
  {Roche}}]{canonico2024orbital}%
  \BibitemOpen
  \bibfield  {author} {\bibinfo {author} {\bibfnamefont {L.~M.}\ \bibnamefont
  {Canonico}}, \bibinfo {author} {\bibfnamefont {J.~H.}\ \bibnamefont
  {Garcia}}, \ and\ \bibinfo {author} {\bibfnamefont {S.}~\bibnamefont
  {Roche}},\ }\href@noop {} {\bibfield  {journal} {\bibinfo  {journal}
  {Physical Review B}\ }\textbf {\bibinfo {volume} {110}},\ \bibinfo {pages}
  {L140201} (\bibinfo {year} {2024})}\BibitemShut {NoStop}%
\bibitem [{\citenamefont {Fan}\ \emph {et~al.}(2021)\citenamefont {Fan},
  \citenamefont {Garcia}, \citenamefont {Cummings}, \citenamefont
  {Barrios-Vargas}, \citenamefont {Panhans}, \citenamefont {Harju},
  \citenamefont {Ortmann},\ and\ \citenamefont {Roche}}]{Fan2021linear}%
  \BibitemOpen
  \bibfield  {author} {\bibinfo {author} {\bibfnamefont {Z.}~\bibnamefont
  {Fan}}, \bibinfo {author} {\bibfnamefont {J.~H.}\ \bibnamefont {Garcia}},
  \bibinfo {author} {\bibfnamefont {A.~W.}\ \bibnamefont {Cummings}}, \bibinfo
  {author} {\bibfnamefont {J.~E.}\ \bibnamefont {Barrios-Vargas}}, \bibinfo
  {author} {\bibfnamefont {M.}~\bibnamefont {Panhans}}, \bibinfo {author}
  {\bibfnamefont {A.}~\bibnamefont {Harju}}, \bibinfo {author} {\bibfnamefont
  {F.}~\bibnamefont {Ortmann}}, \ and\ \bibinfo {author} {\bibfnamefont
  {S.}~\bibnamefont {Roche}},\ }\href {\doibase
  https://doi.org/10.1016/j.physrep.2020.12.001} {\bibfield  {journal}
  {\bibinfo  {journal} {Physics Reports}\ }\textbf {\bibinfo {volume} {903}},\
  \bibinfo {pages} {1} (\bibinfo {year} {2021})}\BibitemShut {NoStop}%
\bibitem [{\citenamefont {Bastin}\ \emph {et~al.}(1971)\citenamefont {Bastin},
  \citenamefont {Lewiner}, \citenamefont {Betbeder-Matibet},\ and\
  \citenamefont {Nozieres}}]{bastin1971quantum}%
  \BibitemOpen
  \bibfield  {author} {\bibinfo {author} {\bibfnamefont {A.}~\bibnamefont
  {Bastin}}, \bibinfo {author} {\bibfnamefont {C.}~\bibnamefont {Lewiner}},
  \bibinfo {author} {\bibfnamefont {O.}~\bibnamefont {Betbeder-Matibet}}, \
  and\ \bibinfo {author} {\bibfnamefont {P.}~\bibnamefont {Nozieres}},\
  }\href@noop {} {\bibfield  {journal} {\bibinfo  {journal} {Journal of Physics
  and Chemistry of Solids}\ }\textbf {\bibinfo {volume} {32}},\ \bibinfo
  {pages} {1811} (\bibinfo {year} {1971})}\BibitemShut {NoStop}%
\bibitem [{\citenamefont {Canonico}\ \emph {et~al.}(2019)\citenamefont
  {Canonico}, \citenamefont {Rappoport},\ and\ \citenamefont
  {Muniz}}]{Luis-PRL}%
  \BibitemOpen
  \bibfield  {author} {\bibinfo {author} {\bibfnamefont {L.~M.}\ \bibnamefont
  {Canonico}}, \bibinfo {author} {\bibfnamefont {T.~G.}\ \bibnamefont
  {Rappoport}}, \ and\ \bibinfo {author} {\bibfnamefont {R.~B.}\ \bibnamefont
  {Muniz}},\ }\href {\doibase 10.1103/PhysRevLett.122.196601} {\bibfield
  {journal} {\bibinfo  {journal} {Phys. Rev. Lett.}\ }\textbf {\bibinfo
  {volume} {122}},\ \bibinfo {pages} {196601} (\bibinfo {year}
  {2019})}\BibitemShut {NoStop}%
\bibitem [{\citenamefont {Jo{\~{a}}o}\ \emph {et~al.}(2020)\citenamefont
  {Jo{\~{a}}o}, \citenamefont {An{\dj}elkovi{\'{c}}}, \citenamefont {Covaci},
  \citenamefont {Rappoport}, \citenamefont {Lopes},\ and\ \citenamefont
  {Ferreira}}]{kite}%
  \BibitemOpen
  \bibfield  {author} {\bibinfo {author} {\bibfnamefont {S.~M.}\ \bibnamefont
  {Jo{\~{a}}o}}, \bibinfo {author} {\bibfnamefont {M.}~\bibnamefont
  {An{\dj}elkovi{\'{c}}}}, \bibinfo {author} {\bibfnamefont {L.}~\bibnamefont
  {Covaci}}, \bibinfo {author} {\bibfnamefont {T.~G.}\ \bibnamefont
  {Rappoport}}, \bibinfo {author} {\bibfnamefont {J.~M. V.~P.}\ \bibnamefont
  {Lopes}}, \ and\ \bibinfo {author} {\bibfnamefont {A.}~\bibnamefont
  {Ferreira}},\ }\href {\doibase 10.1098/rsos.191809} {\bibfield  {journal}
  {\bibinfo  {journal} {Royal Society Open Science}\ }\textbf {\bibinfo
  {volume} {7}},\ \bibinfo {pages} {191809} (\bibinfo {year}
  {2020})}\BibitemShut {NoStop}%
\bibitem [{\citenamefont {Pezo}\ \emph {et~al.}(2022)\citenamefont {Pezo},
  \citenamefont {Garc\'{\i}a~Ovalle},\ and\ \citenamefont
  {Manchon}}]{pezo2022orbital}%
  \BibitemOpen
  \bibfield  {author} {\bibinfo {author} {\bibfnamefont {A.}~\bibnamefont
  {Pezo}}, \bibinfo {author} {\bibfnamefont {D.}~\bibnamefont
  {Garc\'{\i}a~Ovalle}}, \ and\ \bibinfo {author} {\bibfnamefont
  {A.}~\bibnamefont {Manchon}},\ }\href {\doibase 10.1103/PhysRevB.106.104414}
  {\bibfield  {journal} {\bibinfo  {journal} {Phys. Rev. B}\ }\textbf {\bibinfo
  {volume} {106}},\ \bibinfo {pages} {104414} (\bibinfo {year}
  {2022})}\BibitemShut {NoStop}%
\bibitem [{\citenamefont {Liu}\ \emph {et~al.}(2025)\citenamefont {Liu},
  \citenamefont {Cullen}, \citenamefont {Arovas},\ and\ \citenamefont
  {Culcer}}]{liu2025quantum}%
  \BibitemOpen
  \bibfield  {author} {\bibinfo {author} {\bibfnamefont {H.}~\bibnamefont
  {Liu}}, \bibinfo {author} {\bibfnamefont {J.~H.}\ \bibnamefont {Cullen}},
  \bibinfo {author} {\bibfnamefont {D.~P.}\ \bibnamefont {Arovas}}, \ and\
  \bibinfo {author} {\bibfnamefont {D.}~\bibnamefont {Culcer}},\ }\href@noop {}
  {\bibfield  {journal} {\bibinfo  {journal} {Physical Review Letters}\
  }\textbf {\bibinfo {volume} {134}},\ \bibinfo {pages} {036304} (\bibinfo
  {year} {2025})}\BibitemShut {NoStop}%
\bibitem [{\citenamefont {Cullen}\ \emph {et~al.}()\citenamefont {Cullen},
  \citenamefont {Arovas}, \citenamefont {Raimondi},\ and\ \citenamefont
  {Culcer}}]{cullen2025quantum}%
  \BibitemOpen
  \bibfield  {author} {\bibinfo {author} {\bibfnamefont {J.~H.}\ \bibnamefont
  {Cullen}}, \bibinfo {author} {\bibfnamefont {D.~P.}\ \bibnamefont {Arovas}},
  \bibinfo {author} {\bibfnamefont {R.}~\bibnamefont {Raimondi}}, \ and\
  \bibinfo {author} {\bibfnamefont {D.}~\bibnamefont {Culcer}},\ }\href@noop {}
  {\bibinfo  {journal} {arXiv:2505.02911}\ }\BibitemShut {NoStop}%
\bibitem [{sup()}]{suppmaterial}%
  \BibitemOpen
\bibfield  {journal} {  }\href@noop {} {\ }\bibinfo {note} {See Supplemental
  Material at [URL will be inserted by publisher] for a detailed description of
  the KPM, expansion of the OAM operator, spin and orbital characters of the
  $1T_{d}$ TMDs, tight-binding model for the $1T_{d}$ TMD/FM, real-space
  projeciton scheme of the spin and OAM, orbital responses of the isolated
  square FM, and sea and surface contributions of the spin and orbital
  responses of the TMD/FM heterostructure.}\BibitemShut {Stop}%
\bibitem [{\citenamefont {Vila}\ \emph {et~al.}(2021)\citenamefont {Vila},
  \citenamefont {Hsu}, \citenamefont {Garcia}, \citenamefont {Ben{\'\i}tez},
  \citenamefont {Waintal}, \citenamefont {Valenzuela}, \citenamefont
  {Pereira},\ and\ \citenamefont {Roche}}]{vila2021low}%
  \BibitemOpen
  \bibfield  {author} {\bibinfo {author} {\bibfnamefont {M.}~\bibnamefont
  {Vila}}, \bibinfo {author} {\bibfnamefont {C.-H.}\ \bibnamefont {Hsu}},
  \bibinfo {author} {\bibfnamefont {J.~H.}\ \bibnamefont {Garcia}}, \bibinfo
  {author} {\bibfnamefont {L.~A.}\ \bibnamefont {Ben{\'\i}tez}}, \bibinfo
  {author} {\bibfnamefont {X.}~\bibnamefont {Waintal}}, \bibinfo {author}
  {\bibfnamefont {S.~O.}\ \bibnamefont {Valenzuela}}, \bibinfo {author}
  {\bibfnamefont {V.~M.}\ \bibnamefont {Pereira}}, \ and\ \bibinfo {author}
  {\bibfnamefont {S.}~\bibnamefont {Roche}},\ }\href@noop {} {\bibfield
  {journal} {\bibinfo  {journal} {Physical Review Research}\ }\textbf {\bibinfo
  {volume} {3}},\ \bibinfo {pages} {043230} (\bibinfo {year}
  {2021})}\BibitemShut {NoStop}%
\bibitem [{\citenamefont {Xu}\ \emph {et~al.}(2018)\citenamefont {Xu},
  \citenamefont {Ma}, \citenamefont {Shen}, \citenamefont {Fatemi},
  \citenamefont {Wu}, \citenamefont {Chang}, \citenamefont {Chang},
  \citenamefont {Valdivia}, \citenamefont {Chan}, \citenamefont {Gibson} \emph
  {et~al.}}]{xu2018electrically}%
  \BibitemOpen
  \bibfield  {author} {\bibinfo {author} {\bibfnamefont {S.-Y.}\ \bibnamefont
  {Xu}}, \bibinfo {author} {\bibfnamefont {Q.}~\bibnamefont {Ma}}, \bibinfo
  {author} {\bibfnamefont {H.}~\bibnamefont {Shen}}, \bibinfo {author}
  {\bibfnamefont {V.}~\bibnamefont {Fatemi}}, \bibinfo {author} {\bibfnamefont
  {S.}~\bibnamefont {Wu}}, \bibinfo {author} {\bibfnamefont {T.-R.}\
  \bibnamefont {Chang}}, \bibinfo {author} {\bibfnamefont {G.}~\bibnamefont
  {Chang}}, \bibinfo {author} {\bibfnamefont {A.~M.~M.}\ \bibnamefont
  {Valdivia}}, \bibinfo {author} {\bibfnamefont {C.-K.}\ \bibnamefont {Chan}},
  \bibinfo {author} {\bibfnamefont {Q.~D.}\ \bibnamefont {Gibson}},  \emph
  {et~al.},\ }\href@noop {} {\bibfield  {journal} {\bibinfo  {journal} {Nature
  Physics}\ }\textbf {\bibinfo {volume} {14}},\ \bibinfo {pages} {900}
  (\bibinfo {year} {2018})}\BibitemShut {NoStop}%
\bibitem [{\citenamefont {Furukawa}\ \emph {et~al.}(2021)\citenamefont
  {Furukawa}, \citenamefont {Watanabe}, \citenamefont {Ogasawara},
  \citenamefont {Kobayashi},\ and\ \citenamefont {Itou}}]{KMETellurium}%
  \BibitemOpen
  \bibfield  {author} {\bibinfo {author} {\bibfnamefont {T.}~\bibnamefont
  {Furukawa}}, \bibinfo {author} {\bibfnamefont {Y.}~\bibnamefont {Watanabe}},
  \bibinfo {author} {\bibfnamefont {N.}~\bibnamefont {Ogasawara}}, \bibinfo
  {author} {\bibfnamefont {K.}~\bibnamefont {Kobayashi}}, \ and\ \bibinfo
  {author} {\bibfnamefont {T.}~\bibnamefont {Itou}},\ }\href {\doibase
  10.1103/PhysRevResearch.3.023111} {\bibfield  {journal} {\bibinfo  {journal}
  {Phys. Rev. Res.}\ }\textbf {\bibinfo {volume} {3}},\ \bibinfo {pages}
  {023111} (\bibinfo {year} {2021})}\BibitemShut {NoStop}%
\bibitem [{\citenamefont {Go}\ \emph {et~al.}(2018)\citenamefont {Go},
  \citenamefont {Jo}, \citenamefont {Kim},\ and\ \citenamefont
  {Lee}}]{OrbitalTexture}%
  \BibitemOpen
  \bibfield  {author} {\bibinfo {author} {\bibfnamefont {D.}~\bibnamefont
  {Go}}, \bibinfo {author} {\bibfnamefont {D.}~\bibnamefont {Jo}}, \bibinfo
  {author} {\bibfnamefont {C.}~\bibnamefont {Kim}}, \ and\ \bibinfo {author}
  {\bibfnamefont {H.-W.}\ \bibnamefont {Lee}},\ }\href {\doibase
  10.1103/PhysRevLett.121.086602} {\bibfield  {journal} {\bibinfo  {journal}
  {Phys. Rev. Lett.}\ }\textbf {\bibinfo {volume} {121}},\ \bibinfo {pages}
  {086602} (\bibinfo {year} {2018})}\BibitemShut {NoStop}%
\bibitem [{\citenamefont {Medina~Due{\~n}as}\ \emph {et~al.}(2024)\citenamefont
  {Medina~Due{\~n}as}, \citenamefont {Garc{\'\i}a},\ and\ \citenamefont
  {Roche}}]{medina2024emerging}%
  \BibitemOpen
  \bibfield  {author} {\bibinfo {author} {\bibfnamefont {J.}~\bibnamefont
  {Medina~Due{\~n}as}}, \bibinfo {author} {\bibfnamefont {J.~H.}\ \bibnamefont
  {Garc{\'\i}a}}, \ and\ \bibinfo {author} {\bibfnamefont {S.}~\bibnamefont
  {Roche}},\ }\href@noop {} {\bibfield  {journal} {\bibinfo  {journal}
  {Physical Review Letters}\ }\textbf {\bibinfo {volume} {132}},\ \bibinfo
  {pages} {266301} (\bibinfo {year} {2024})}\BibitemShut {NoStop}%
\end{thebibliography}
\end{document}